\documentclass[12pt]{iopart}

\usepackage{amssymb}
\usepackage{graphicx}
\usepackage{color}
\usepackage{caption}
\usepackage{soul}

\newcommand{\grafe}[1]{\left\{ #1 \right\}}
\newcommand{\tonde}[1]{\left( #1 \right)}
\newcommand{\quadre}[1]{\left[ #1 \right]}

\newcommand{\abs}[1]{\left| {#1} \right|}

\definecolor{darkGreen}{RGB}{0,110,0}
\definecolor{darkBlue}{RGB}{0,0,130}

\newcommand{\bra}[1]{\left\langle #1 \right|}
\newcommand{\ket}[1]{\left| #1 \right\rangle}

\usepackage{array}
\newcolumntype{L}{>{$}l<{$}}

\begin{document}

\title{Anderson transition on the Bethe lattice: an approach with real energies}
\author{Giorgio Parisi$^{1,2}$, Saverio Pascazio$^{3,4}$, Francesca Pietracaprina$^{1,5}$, Valentina Ros$^{6}$ and Antonello Scardicchio$^{7,8}$ }
\address{$^1$Dipartimento di Fisica, Universit\`{a} di Roma ``La Sapienza'', Piazzale Aldo Moro 2, I-00185 Roma, Italy}
\address{$^2$Centre for Statistical Mechanics and Complexity (SMC), CNR-INFM, I-00185 Roma, Italy; INFN, Sezione di Roma,  I-00185 Roma, Italy}
\ead{giorgio.parisi@roma1.infn.it }
\address{$^3$Dipartimento di Fisica and MECENAS, Universit\`{a} di Bari, I-70126 Bari, Italy}
\address{$^4$ INFN, Sezione di Bari, I-70126 Bari, Italy;
Istituto Nazionale di Ottica (INO-CNR), I-50125 Firenze, Italy}
\ead{saverio.pascazio@ba.infn.it}
\address{$^5$ Laboratoire de Physique Th\'eorique, IRSAMC, Universit\'e de Toulouse, CNRS, UPS, France}
\ead{pietracaprina@irsamc.ups-tlse.fr}
\address{$^6$Laboratoire de Physique de l'\'Ecole normale sup\'erieure, ENS, Universit\'e PSL, CNRS, Sorbonne Universit\'e, Universit\'e Paris-Diderot, Sorbonne Paris Cit\'e, Paris, France}
\ead{valentina.ros@lps.ens.fr}
\address{$^7$The Abdus Salam International Center for Theoretical Physics, Strada  Costiera  11,  34151  Trieste,  Italy}
\address{$^8$ INFN Sezione di Trieste, Via Valerio 2, 34127 Trieste, Italy}
\ead{ascardic@ictp.it }

\vspace{10pt}
\begin{indented}
\item[]May 2019
\end{indented}
\begin{abstract}
We study the Anderson model on the Bethe lattice by working directly with propagators at real energies $E$. We introduce a novel criterion for the localization-delocalization transition based on the stability of the population of the propagators, and show that it is consistent with the one obtained through the study of the imaginary part of the self-energy. We present an accurate numerical estimate of the transition point, as well as a concise proof of the asymptotic formula for the critical disorder on lattices of large connectivity, as given in Ref. \cite{Anderson1958}. We discuss how the forward approximation used in analytic treatments of localization problems fits into this scenario and how one can interpolate between it and the correct asymptotic analysis.
\end{abstract}

%
%
%
%
%

\section{Introduction}

Anderson Localization \cite{Anderson1958,ThoulessReview} is one of the deepest and most universal quantum mechanical implications for the dynamics of disordered systems. Under certain circumstances (small spatial dimension, large disorder) quantum mechanical effects can be so strong as to completely hinder transport in models of non-interacting particles hopping on lattices (the so-called Anderson models). When this happens, one speaks of an Anderson localized phase. Although the localized phase has come under almost complete mathematical control \cite{stolz2011introduction,aizenman2011resonant}, the complementary (\textit{``delocalized"}) phase, and in particular the phase transition between them has proved elusive to most analytic control, despite the constantly improving abundance and quality of numerical results \cite{Evers2008Review}. Some exceptions to this is provided by the $\epsilon$ expansion of Efetov \cite{efetov1983supersymmetry}, for the Anderson model on a $d$-dimensional square lattice $\mathbb{Z}^d$ around $d=2$. However, these results fail to predict the behavior of the critical point for $d\gtrsim 3$ (see for example \cite{tarquini2017critical}).

Following the similarity with conventional thermodynamic transitions, one is then led to consider loop-less lattices as proxies for the $d\to\infty$ limit. The simplest example of loop-less geometry is probably the Bethe lattice and several works, starting with \cite{abou1973selfconsistent} and including \cite{mirlin1991localization}, have dealt with the Anderson model on the Bethe lattice. The Bethe lattice, as the infinite limit of a regular random graph \cite{bollobas1998random}, might look like a peculiar geometry to study a nearest-neighbor hopping problem, but it is quite common in spin glasses models for two apparently disconnected reasons: on one hand, due to absence of short loops in the lattice \cite{mezard1987spin,mezard2001bethe}, one can make analytic progress in writing a mean field theory {in terms of recursion equations, see Ref. \cite{abou1973selfconsistent}};  on the other hand this lattice structure naturally arises in random optimization problems \cite{monasson1999determining} and therefore the results have great relevance for applications \cite{mezard2002analytic}.
For Anderson Localization (AL) {a further reason that is behind the recent renaissance of interest in the problem lies in the connection, first emphasized in Ref.~\cite{altshuler1997quasiparticle}, between AL on the Bethe lattice and localization effects in interacting many-body systems, a topic now known as Many-Body Localization (MBL)\cite{Fleishman1980interactions,Basko:2006hh}}. MBL has been at the center of a prolific research activity \cite{gornyi2005interacting, oganesyan2007localization,vosk2013dynamical,nandkishore2015many,abanin2017recent}, which, in addition to the absence of transport (a trait in common with AL), has uncovered some peculiarities like the behavior of entanglement measures \cite{PhysRevB.77.064426,Serbyn:2013he,John2015TotalCorrelations}, emergent integrability \cite{huse2013phenomenology,serbyn2013local,ros2015integrals,imbrie2017review}, and protection of symmetries at high temperature~\cite{huse2013localization}. In a recent line of research, some of the MBL phenomenology has been found in highly excited states of systems without disorder~\cite{Papic2015,Schiulaz2015,Hickey2016} and even lattice gauge theories~\cite{PhysRevLett.120.030601}.

Pushing ahead on the initial motivation for connecting the Anderson model on the Bethe lattice with MBL (where regions of {\it sub-diffusing transport} are seen to exist within the ergodic phase \cite{Reichman2014Absence,Luitz2016Extended,Znidaric2016Diffusive,luitz2017ergodic,schulz2018energy,mendoza2018asymmetry}), a conjecture mostly based on numerical evidence has been put forward \cite{de2014anderson} (see also the previous works \cite{altshuler1997quasiparticle, Biroli:2012vk}), and consequently made into a phenomenology \cite{altshuler2016nonergodic,kravtsov2017non}. According to these papers, the eigenstates of the Anderson model on \emph{finite} $N$-vertices Regular Random Graphs (RRG) are non-ergodic, in that they possess anomalous dimensions, in a way similar to what is found in models of random matrices  \cite{kravtsov2015random, Monthus2017Matrices}. More recently, a corresponding putative dynamical transition in the return probability \cite{bera2018return} has been observed. This picture has been challenged by other numerical~\cite{tikhonov2016anderson, garcia2017scaling} and analytical studies~\cite{TikhMirlin2018, MetzLevelCompressibility}, that support the scenario according to which ergodicity is restored for system-sizes exceeding a certain critical volume.
This issue can be roughly summarized into the question: ``are all the properties of the eigenfunctions and of the dynamics encoded in the recursion equations introduced in Ref.~\cite{abou1973selfconsistent}?''
While the question of ergodicity or lack thereof concerns finite lattices, and in particular regular random graphs with $N$ vertices having plenty of loops of length $O(\ln N)$, we feel that a more detailed numerical and analytical knowledge of the recursion relations in Ref.~\cite{abou1973selfconsistent} is welcome, at least to dispel claims that some of the features observed in the RRG numerics are directly implied by the recursion equations.

{ In this work we provide a new perspective on the recursion relations and derive implications on the above and related issues.  Our strategy will be to work with propagators computed  directly at \emph{real} values of the energy, at variance with the more conventional approaches in which the energies are treated as complex variables. This choice makes the numerics more convenient and more readily comparable to the exact diagonalization results, while the analytical calculations are not more complicated than in the conventional approach.}

{ The organization of the article is as follows: in Section~\ref{sec:RealAmpl0} we discuss how the Anderson transition can be described in terms of the susceptibilities of the populations of propagators, without introducing an imaginary part for the energy. In Section~\ref{sec:inteq} we show how the integral equation introduced in \cite{abou1973selfconsistent} to determine the stability of the Anderson phase can be recovered from the probability distribution of the susceptibilities. We discuss some features of its solution for two different distributions of the disorder: the uniform distribution (conventionally considered in the numerics) and the Cauchy distribution, for which analytic results can be derived. Using the formalism based on the integral equation, we derive analytically the weak-disorder behaviour of the fractal dimension  $D_1$ defined as in Ref.~\cite{kravtsov2017non}. In Section \ref{sec:largedis} we revisit the large-connectivity limit of the problem, bringing up a new point of view on the integral equation. We present a concise proof of the known asymptotic formula \cite{Anderson1958, abou1973selfconsistent, Bapst2014HighConnectivity} for the critical disorder $W_c$ at large connectivity, which relies on the identification of the eigenvector corresponding to the largest eigenvalue of the integral kernel. We then discuss why the asymptotic formula is not correctly recovered within the forward or ``upper limit'' approximation of Ref.~\cite{Anderson1958}, and introduce a generalized integral equation that interpolates between the forward approximation \cite{pc2016forward}  and to the full Anderson problem at large connectivity.}

\section{{Real-energy propagators:} the criterion for localization}\label{sec:RealAmpl0}

\subsection{On the inverted thermodynamic limit}
\label{sec:RealAmpl}

We consider the Anderson model on a Random Regular Graph with connectivity $Z=K+1$ ($K$ being the branching number).  In the infinite volume limit, any finite section of this graph becomes tree-like. The Anderson model has the Hamiltonian:
\begin{equation}
    H= \sum_i \epsilon_i |i \rangle \langle i| + t\sum_{\langle i,j \rangle} \tonde{| i \rangle \langle j|+ |j \rangle \langle i|},
\end{equation}
where $i$ labels the vertices of the graph, the second sum is over nearest neighboring vertices, and $\epsilon_i$ are identically distributed independent random variables. We mostly focus on two distributions of disorder $\rho(\epsilon)$: a uniform distribution with $\epsilon_i\in[-W/2,W/2]$, and the Cauchy distribution $\rho(\epsilon_i)=(W/\pi)/(W^2+\epsilon_i^2).$ {The latter is particularly interesting as it allows to derive analytic results, mainly due to its stability with respect to the iteration of the recursion relations. We notice that the ubiquity of the Cauchy distribution for random Schr\"odinger operators has been discussed in the mathematical physics literature in the past \cite{aizenman2011resonant} and cannot possibly be overemphasized.}

In the following, we set $t=1$ for simplicity, and we focus on the case $E=0$, $E$ being the energy, unless stated otherwise. In the numerics we always consider $K=2$.
The diagonal part of the resolvent, \emph{i.e.}, the Fourier transformed survival amplitude:
\begin{equation}
\mathcal{G}_{ii}(E)=\bra{i}\frac{1}{E-H}\ket{i},
\label{eq:Gii}
\end{equation}
can be computed from the approximated equations~\cite{abou1973selfconsistent}
\begin{equation}
\mathcal{G}_{ii}(E)=\frac{1}{E-\epsilon_i-\sum_{k=1}^{K+1}G_{k,k}(E)},
\label{eq:GiiGk}
\end{equation}
where the sum is over all sites $k$ that are nearest neighbors of $i$, and the $G_{k,k}(E)$'s are {\it cavity Green functions}: they correspond to the Hamiltonian restricted to the sub-tree rooted in site $k$, once the link connecting sites $i$ and $k$ is removed. These equations are expected to become exact in the infinite volume limit. The exact relation, valid for any size and any graph, is of the form:
\begin{equation}
\mathcal{G}_{ii}(E)=\frac{1}{E-\epsilon_i-\sum_{k=1}^{K+1}\sum_{j=1}^{K+1}G_{k,j}(E)},
\label{eq:GiiGkbis}
\end{equation}
which reduces to the previous equation if the off-diagonal terms $G_{k,j}$ with $k\neq j$ are negligible for large $N$.

Why should the off-diagonal terms be negligible? We first note that the points $k$ and $j$, in absence of $i$, are not directly connected: they are not connected at all on a tree geometry while they are at distance of the order of $\ln(N)/\ln(K)$ on a RRG. We also notice that
\begin{equation}
G_{k,j}=\sum_{\alpha}{\psi^\alpha_k\psi^\alpha_j\over E-E^\alpha},
\end{equation}
where the $\psi^\alpha_k$ and the $E^\alpha$ are the eigenvectors and the eigenvalues of the cavity Hamiltonian.
The typical values of the quantities $\psi^\alpha_k\psi^\alpha_j$ are of order $1/N$ in the case of ergodic states, and are exponentially small in the distance between $j$ and $k$ in the case of localized states.
When $E$ has an imaginary part the off-diagonal terms are always small. The same happens for real $E$ and finite $N$, where problems may arise when the value of $E$ belongs to the support of the spectrum, as the denominators vanish. In the case of localized states the off-diagonal terms are small; also in the case of extended non-ergodic states the off-diagonal terms can be neglected, while they have to be considered in the case of extended ergodic states.
Under the hypothesis that the diagonal terms can be neglected, by following the usual cavity construction one can easily show that the cavity Green functions at energy $E$ satisfy the recursion
\begin{equation}
G_{i}(E)=\frac{t^2}{E-\epsilon_i-\sum_{k=1}^K G_k(E)},
\label{eq:Gcav}
\end{equation}
where the sum is over the $K$ cavity neighbors of site $i$ and we introduced the shorthand notation $G_k=G_{k,k}$ for the diagonal terms. These equations are defined here for a finite tree geometry; the usual route to analyze them consists in adding a small imaginary part to $E\to E+i\eta$, which is sent to zero after the thermodynamic limit is taken. In the following we shall follow an alternative route and consider $\eta \to 0$.
The inversion of the limits $\eta \to 0$ and $N\to\infty$, dubbed ``inverted thermodynamic limit" in Ref.~\cite{kravtsov2017non}, has been considered in \cite{kravtsov2017non, PhysRevB.94.184203, mirlin2000statistics} to deal with the eigenfunction statistics. It is clear that in the inverted thermodynamic limit the $G$'s are always real; in the standard thermodynamic limit, in the extended case the equation admits one solution where the $G$'s are real, that is however unstable with respect to the addition of imaginary parts.

The use of the simplified cavity equation without the off-diagonal terms may look unfounded, as far as it is easy to convince ourselves that the off-diagonal terms cannot be neglected in the extended ergodic phase. Indeed, the equations cannot be used in this phase unless one proves that the extended states are non ergodic. Of course we can (and will) prove that these equations can be used to show that the extended states are not ergodic.
Before proceeding, we have to face a serious consistency problem. In the standard thermodynamic limit in the extended non-ergodic case we have a probability distribution of a complex $G=\hat R+i\hat I$ that satisfies the same standard cavity equation in the inverted thermodynamic limit with a probability distribution of the real $G$. At the end of the day we are left with two probability distributions:
\begin{equation}
P(\hat R,\hat I) \quad \mbox{    and    }\quad P(G)\,.
\end{equation}
Both probability distributions describe the physics of the problem. The consistency problem becomes more acute if we notice that the spectral density $\rho(E)$ can be extracted by both equations, as:
\begin{equation}
\rho(E)=\int d\hat R \,d\hat I P(\hat R,\hat I) \hat I\  \quad  \mbox{and} \   \lim_{G\to\infty}G^2P(G)=\rho(E).
\end{equation}
This consistency problem has a simple solution: the two probability functions are not independent. Indeed, $P(G)$ can be written as a simple function of $P(\hat R,\hat I)$
\begin{equation}
P(G)= \int d\hat R\, d\hat I P(\hat R,\hat I) {\hat I\over (\hat R-G)^2+\hat I^2}\,. \label{MAGIC}
\end{equation}
This relation was already given in \cite{Miller1994}, and it is quite possible that it has its root in the hyperbolic geometry of the $O(n|n)$ group in the $n\to 0$ limit. The proof is quite direct. It makes use the basic properties of the Cauchy distribution: it is quite simple to check that if $ P(\hat R,\hat I)$ satisfies the {\sl complex} cavity equation, the function $P(G)$ defined in Eq. (\ref{MAGIC}) satisfies the {\sl real} cavity equation. Notice that, as mentioned before, the inconsistency exists only in the extended non-ergodic case: in the localized case the distribution $P(\hat R,\hat I)$ is concentrated on $\hat I=0$, and in the extended ergodic case there is no simple equation for $P(G)$.

Having checked that the results of the inverted thermodynamic limit are consistent with those of the standard thermodynamic limit, we will devote the rest of the paper to the study of this limit.

\subsection{Vanishing disorder: {the emergence of the Cauchy distribution}}
\label{sec:nuova}

\begin{figure}[htbp]
\begin{center}
\includegraphics[width=.495 \columnwidth]{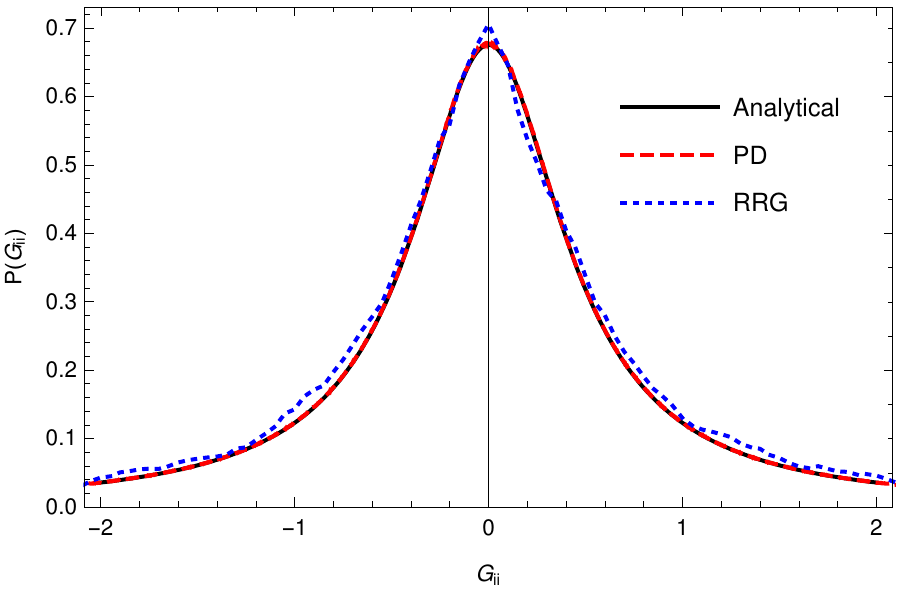}
\includegraphics[width=.495\columnwidth]{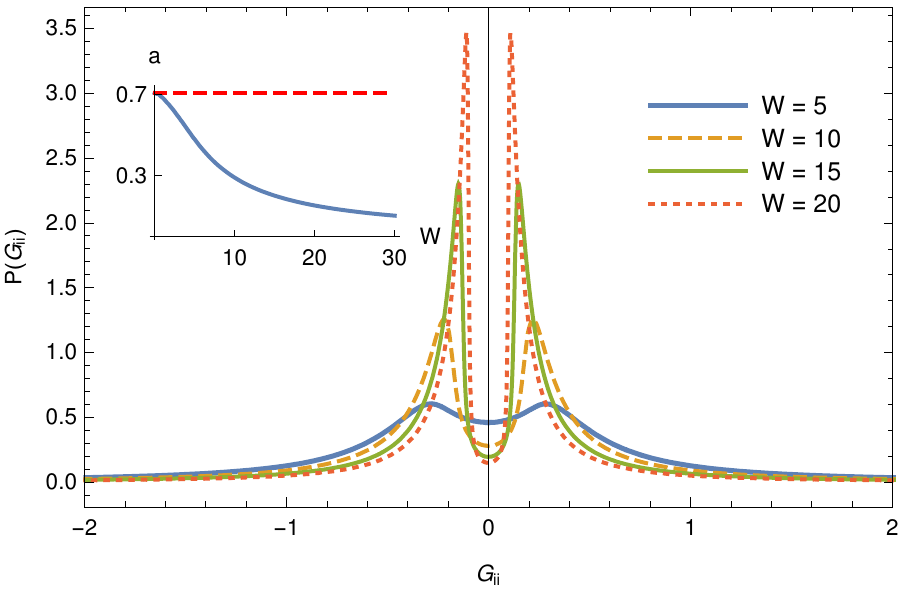}
\caption{{\it Left panel.} Probability distribution of $\mathcal{G}_{ii}$ for $K=2$ and $W=0$. Comparison between the analytical solution, the exact numerical solution (for a RRG with $N=2^{16}$ nodes) and numerical population dynamics.  {\it Right panel.} Distribution of the full propagator $\mathcal{G}_{ii}$ obtained from population dynamics for population size $N = n s = 10^8$ and four values of disorder: $W=5, 10, 15$ and $20$. \textit{ Right panel  - Inset. } Coefficient of the Cauchy-like ($\sim G^{-2}$) tails of $P(G)$.}\label{fig:RRGW0}
\end{center}
\end{figure}

Consider first the limiting case of vanishing disorder $W=0$. Setting $E=0$, the equation for $G_i$ reduces to
\begin{equation}
G_i=\frac{-1}{\sum_{k=1}^K G_k},
\label{eq:cav_rec_clean}
\end{equation}
which has no unique real solution (assuming $G_i=G$). This is natural, since the resolvent is undefined for values of $E$ belonging to the spectrum of the corresponding operator (in this case, the adjacency matrix of the graph). However, the above equations admit a solution in distribution in the form of a Cauchy-Lorentz distribution of width $1/\sqrt{K}$
\begin{equation}
P(G)=\frac{(\pi\sqrt{K})^{-1}}{K^{-1}+G^2}.
\label{eq:PGCclean}
\end{equation}
This, in turn, induces a Cauchy-Lorentz distribution for the $\mathcal{G}_{ii}$
\begin{equation}
\mathcal{P}(\mathcal{G})=\frac{b/\pi}{b^2+\mathcal{G}^2},
\label{eq:PGclean}
\end{equation}
where $b=\sqrt{K}/(K+1)$. This agrees with numerical results, as shown in Figure \ref{fig:RRGW0}, where the distribution of $\mathcal{P}$ in (\ref{eq:PGclean}) is compared with numerical solutions of Eq. (\ref{eq:Gii}) on an ensemble of RRGs. The fact that (for random operators) the diagonal matrix elements of the resolvent evaluated on the real axis are Cauchy variables is a property that holds in surprising generality \cite{aizenman2015ubiquity, PotterShortPieces},  irrespectively of the statistics of the matrix eigenvalues and of the possible strong correlations between them, as exemplified by Coulomb gas models \cite{Brouwer, fyodorov2013universal,PotterShortPieces}. Moreover, the fact that the numerics on RRG is so accurately described by a cavity equation means that the correlations along closed paths of length $\ln_{K}N$ which loop around the RRG are irrelevant for this quantity, and that the $G_{k}$'s can be considered as independent, identically distributed variables. This is due to the presence of a gap in the spectrum of the integral equation associated with the susceptibility, as also shown below.

\subsection{Finite disorder: population dynamics {and exact numerics on RRG}}
\label{sec:PD}

In presence of disorder, the stable probability distribution that describes the random variables $G_i$ satisfies the cavity equation
\begin{eqnarray}
P(G)&=\int d\epsilon \rho(\epsilon)\int d^K G \prod_{i=1}^K P(G_i) \delta\left(G-\frac{1}{E-\epsilon-\sum_k G_k}\right).
\label{eq:PGeq}
\end{eqnarray}
We determine the distribution of the cavity propagators $G_i$ and of $\mathcal{G}_{ii}$ by using a (by now standard) population dynamics algorithm: we iterate Eq. (\ref{eq:Gcav}) over a population of size $n$ (up to $n=10^7$), initially drawn from a uniform distribution. We perform $I=100$ sweeps of the population to achieve convergence, replacing each element of the population (the new elements are thus available to be drawn before the sweep is complete). We then sample the population $s=10$ times at iteration $i=100, 115 \dots 250$. This gives the same result as iterating a population of size $N = n s$ directly.

To compare with exact results on the RRG geometry, we also determine $\mathcal{G}_{ii}$ by numerically solving the linear equation
\begin{eqnarray}
\ket{i}=(H-E)\sum_{j}\mathcal{G}^{RRG}_{ij}\ket{j},
\end{eqnarray}
using for the kinetic term in $H$ the adjacency matrix of the regular random graph (with hopping $t=1$), which takes loops into consideration.
The above is the equation for a vector $v=\sum_j\mathcal{G}^{RRG}_{ij}\ket{j}$ in the canonical form $v_0=M v,$ where $v_0$ and $M$ are known. Since linear solvers are more efficient than exact-diagonalization algorithms, a lot of statistics can be accumulated in a short time.

\begin{figure}[htbp]
\begin{center}
\includegraphics[width=.495 \columnwidth]{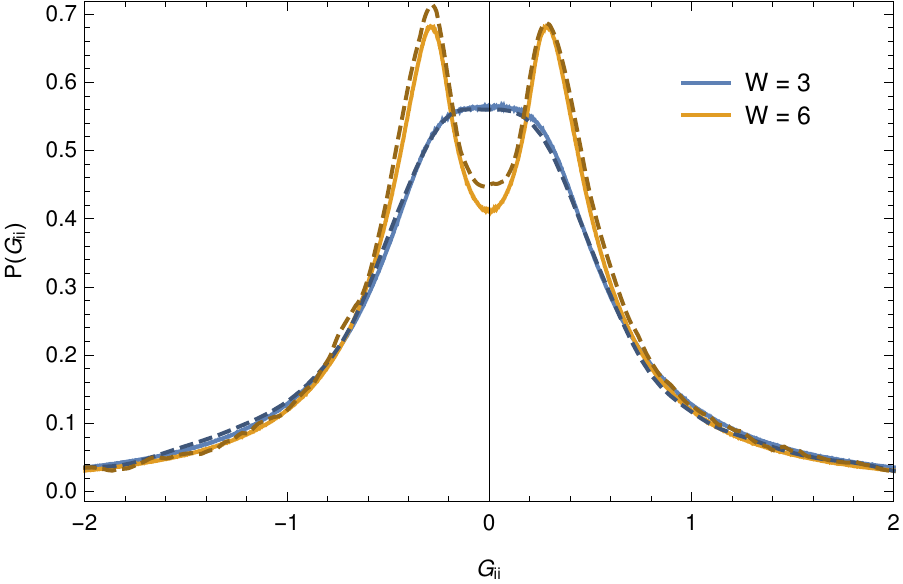}
\includegraphics[width=.495\columnwidth]{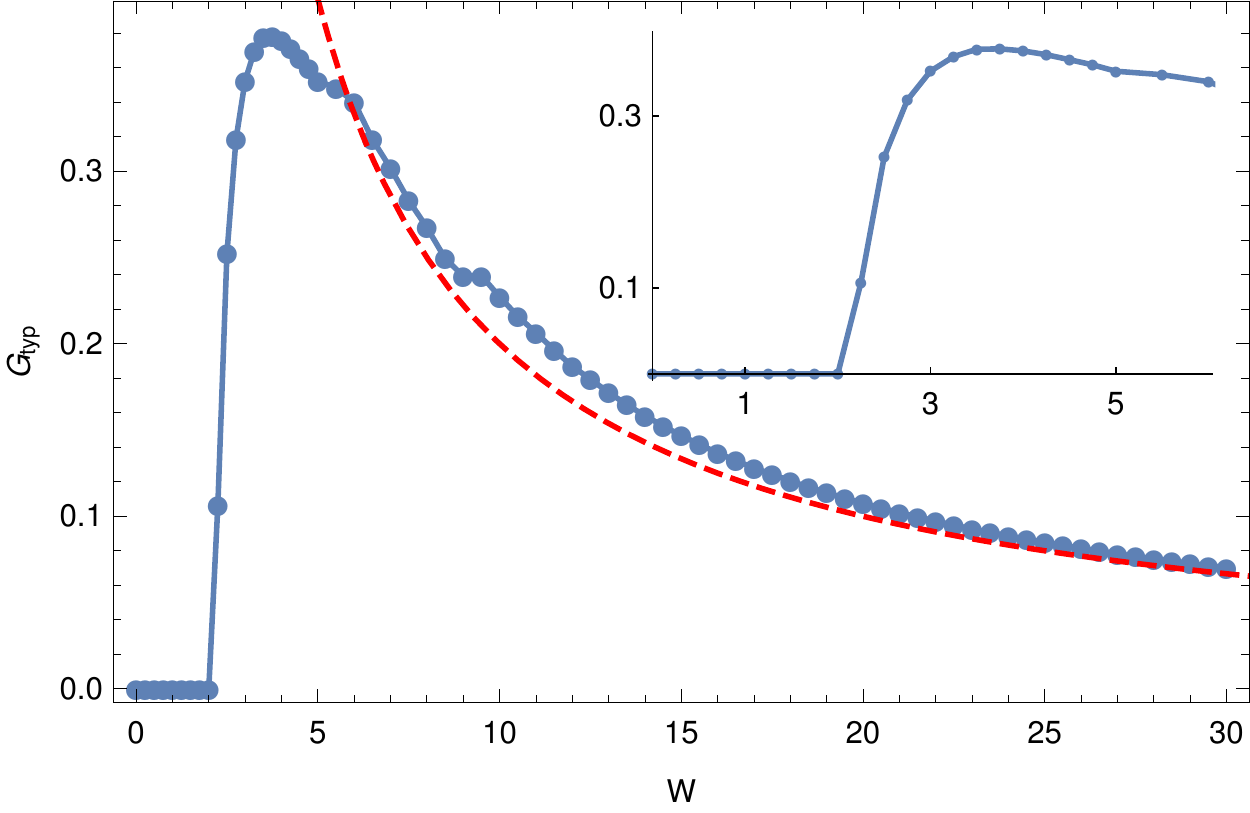}
\caption{{\it Left panel.} Comparison of the exact numerical solution in a RRG with $N=2^{16}$ nodes (dashed line) and population dynamics (solid line) for the probability distribution of $\mathcal{G}_{ii}$, for $K=2$ and disorder $W=3$ and $W=6$. {\it Right panel.} The most probable value of $G$ (maximum of $P(G)$) as a function of disorder. The dashed line is the estimate $2/W$, valid at large disorder.  \textit{ Right panel - Inset. } Zoom of the small-disorder region.}\label{fig:RRGW}
\end{center}
\end{figure}

The distribution of $\mathcal{G}^{RRG}_{ii}$ and the one of $\mathcal{G}_{ii}$ obtained from the analysis of the population dynamics agree within the precision of the numerical computations. For $W>0$, the distribution $P(G)$ for boxed disorder shows a depletion around $G=0$ (see Fig.~\ref{fig:RRGW0}, as well as the comparison with the exact numerical RRG result in Fig.~\ref{fig:RRGW}).  A consequence of it is that the maximum of $P(G)$ moves from $G=0$ to $G>0$ discontinuously at a certain $W\simeq 2.2$, see Fig.~\ref{fig:RRGW}. For large $W$, as we argue in the following, $P(G)\simeq G^{-2}\rho(1/G)$, giving $P(G) \propto G^{-2}$ for large-enough $G$.  {More precisely, a good large-disorder approximation of P(G) is given by:}
\begin{equation}
P(G)\simeq \frac{1}{1+ 4K/W^2} \left\{ \begin{array}{@{\kern2.5pt}lL}
    \hfill (W G^2)^{-1} & if $|G| \geq 2/W$\\
          K/W & if $|G|<2/W$,
\end{array}\right.
\end{equation}
{where the small-$G$ behaviour is obtained using the tails of the distribution of $G_i$ together with  $G \sim (\sum_{k=1}^K G_i)^{-1}$. This
 yields the location of the maximum at $G=2/W$ for box disorder distribution.} This is a good estimate at sufficiently high disorder (say $W>6$ for $K=2$), as shown in the inset of {Fig. \ref{fig:RRGW}}. The tails of the distribution behave as in Eq. (\ref{eq:PGclean}), i.e. $a/\pi (a^2+G^2)$, for any disorder (with $a \to K^{-1/2}$ as $W \to 0$).  {The coefficient $a$ can be obtained from a direct fit of the tails of $P(G)$; an alternative, more convenient way to extract it numerically is through the distribution $p(l)$ of the quantity $l=(1+W/2 G)^{-1}$, as $a=\lim_{l\to0}p(l)/W$.} The coefficient $a$ can then be used (with proper normalization) to extrapolate the distribution $P(G)$ on the full real axis.

By looking at real populations one cannot directly see if the system is in a localized or delocalized phase. Rather, the populations must be used as input for a more refined analysis, which is done in the next sections.

\subsection{A criterion for the transition {based on susceptibilities}}\label{sec:CritSusc}

When $E=E_r+i\eta$, the localized (delocalized) phase is characterized by the (in)stability of the phase in which $\Im  G$ is concentrated on its limiting value, related to the local density of states.
For real $E$, one can formulate the problem in terms of the \emph{stability of a given population}, or at the decay of the susceptibility. In the localized region, the change of the value of an energy level at position $i$ should not result in a modified dynamics of a site $j$, when the distance between $i$ and $j$, $d(i,j)=L$, goes to infinity (in the sense that $d\gg \xi$, the localization length). Therefore one is led to consider the decay with $L$ of the telescopic identity associated with the shortest unique path $p$ from $i$ to $j$
\begin{equation}\label{eq:susceptP}
\chi_{p}=\frac{\partial G_j}{\partial \epsilon_i}=\left(\prod_{k=1}^L\frac{\partial G_{k+1}}{\partial G_k}\right)\frac{\partial G_i}{\partial \epsilon_i}\propto\prod_{k=0}^L G_k^2,
\end{equation}
where $k$'s are all the sites on the (single) path between $j$ and $i$, where $k=0$ corresponds to the site $i$ and $k=L+1$ to the end site $j$.

If one considers a RRG, the number of paths going from one site to the other is one for $L\lesssim \ln N/\ln K$ but quickly becomes $K^L$
when $L\gtrsim \ln N/\ln K$ (see \cite{bollobas1998random,marinari2004circuits}). The susceptibility is then obtained by summing (\ref{eq:susceptP}) over $K^L$ paths,
\begin{equation}\label{eq:suscept}
\chi_L= {\sum_{p=1}^{K^L} \chi_p \propto} \sum_{p=1}^{K^L}\prod_{k\in p} G_k^2.
\end{equation}
The Lyapunov exponent governing the large-$L$ behavior of the susceptibility is defined by
\begin{eqnarray}
\Lambda=\lim_{L\to\infty}\frac{1}{L}\ln \chi_L.
\label{eq:lambdas}
\end{eqnarray}
While the susceptibility $\chi_L$ is a wildly fluctuating random number, the Lyapunov exponent $\Lambda$ is instead self-averaging. {The system is localized whenever its typical value is negative: the values of parameters at which $\Lambda=0$ correspond to the transition point.}

To get an expression for the Lyapunov exponent $\Lambda$ let us first consider the distribution $Q(\chi)$ of a single path contribution $\chi_p$ (we drop for convenience the index $p$). This can be extracted from its $s$-th moment, interpreted as a Mellin transform of $Q(\chi)$:
\begin{equation}
\langle \chi^s\rangle=\int_0^\infty d\chi Q(\chi)\chi^s=\langle \prod_{k=1}^LG_k^{2s}\rangle=C_L \lambda(s)^L,
\label{eq:chis}
\end{equation}
where we have absorbed the exponential dependence in $\lambda(s)$, and thus assumed that $C_L$ grows less than exponentially in $L$. We introduce a generalized exponent $\mu(s)$ through the equality $\lambda(s)= e^{\mu(s)}$ and invert the Mellin transform, getting
\begin{equation}
Q(\chi)=C_L\int_B\frac{ds}{2\pi i}\chi^{-s-1}e^{L\mu(s)},
\label{eq:Qchi}
\end{equation}
where the integral is done over the Bromwich path, parallel to the imaginary axis.
Using the notation $\theta=(\ln\chi)/L$ we have
\begin{equation}
Q(\chi)=C_L \int_B\frac{ds}{2\pi i}e^{-L[(s+1)\theta-\mu(s)]}\Big|_{L \theta= \log \chi} .
\label{eq:InverseLap}
\end{equation}
Equivalently, the distribution $Q(\theta)$ of the rescaled variable $\theta$ reads
\begin{equation}\label{eq:DistrTheta}
Q(\theta)=C_L \int_B\frac{ds}{2\pi i}e^{-L(s\theta-\mu(s))}.
\end{equation}

For large $L$, the integrals can be computed by the saddle point method, by passing through the real saddle point $s^*$, in a direction parallel to the imaginary axis, where $s^*=s^*(\theta)$ is an implicit function of $\theta$ defined by the condition:
\begin{equation}\label{eq:CondTheta}
\mu'(s^*)=\theta.
\end{equation}
Plugging this back in \eref{eq:DistrTheta} we have
\begin{equation}\label{eq:DistQ}
Q(\theta)\simeq e^{-L(s^*\theta-\mu(s^*))}.
\end{equation}
The value $\theta=0$ discriminates between an exponentially growing ($\theta>0$) and an exponentially decaying ($\theta<0$) susceptibility along a single path. The \emph{typical} value of the path susceptibility is governed by $\theta_{\rm typ}$ satisfying
\begin{equation}\label{eq:ThetaTyp}
\theta_{\rm typ}=\lim_{s\to 0}\frac{\partial \mu(s)}{\partial s},
\end{equation}
implying the vanishing of the large-deviation function for $\theta$, see Eq.\ \eref{eq:DistrTheta}. The transition between the localized and delocalized phases is actually governed by \emph{atypically large} fluctuations of the single-path susceptibilities; the typical decay \eref{eq:ThetaTyp} has however been considered in the literature\footnote{Notice that these references consider the typical value of the squared susceptibility.}: it appears in a sufficient criterion for delocalization in Ref. \cite{Aizenman2013}, and it has been conjectured to be related to the transition between ergodic and non-ergodic phases in the delocalized phase in Ref. \cite{kravtsov2017non}. We discuss this more extensively in the next section.

{A change of variables from $\theta$ to $\chi$ shows that} the distribution of the product $\prod_{k\in p} G_k^2$ has long tails, implying that the sum over paths in \eref{eq:suscept} is dominated by the maximal among the summands. The typical value of the total susceptibility thus solves the equation $K^L Q(\chi_L^{\rm typ})=1$. The transition corresponds to this typical value becoming equal to one (corresponding to $\Lambda=0$), and thus it is obtained setting $\theta=0$ in \eref{eq:DistQ}. In this case, one is left with the equation $\mu'(s^*)=0$ to solve.
As first noted in Ref.\ \cite{abou1973selfconsistent} (and in a more general setting in Ref.\ \cite{altshuler1989distribution}), the symmetry of the equation relating $G_k$ to $G_{k-1}$ implies that  $\mu'(s)=0$ iff $s=1/2$. Keeping this in mind, the criterion for the transition reduces to:
\begin{equation}
\ln K+\mu(s^*)=0,
\label{eq:transitionmu}
\end{equation}
or
\begin{eqnarray}
K\,\lambda(s^*=1/2)=1,
\end{eqnarray}
where $\mu(s)=\ln \lambda(s)$, which is exactly the resonant criterion derived in Ref. \cite{aizenman2011resonant}. This is not surprising, as the total susceptibility appears naturally when linearizing the self-consistent equations for the self energy around the solution having zero imaginary part \cite{aizenman2011resonant}, which is stable in the localized phase.
Close to the critical point, and for $\chi=O(1)$, one can even write the full probability distribution of $\chi$ by noticing that, close to $\theta=0$, by writing $s^*=1/2+\epsilon$ we have $\epsilon=\theta/\mu''(1/2)$ and so
\begin{equation}
Q(\chi) =e^{-L(\ln K+\frac{3}{2}\theta+\frac{\theta^2}{2\mu''(1/2)})}=\frac{K^{-L}}{\chi^{3/2}}e^{-\frac{(\ln\chi)^2}{2L\mu''(1/2)}}
\label{eq:Qchi_2}
\end{equation}
is a log-normal distribution.

There are two ways of tackling the calculation of $\lambda(s)$: one can directly compute the distance between two populations initially differing by a small quantity, or one can link $\lambda(s)$ to the eigenvalue of an integral kernel, {mapping explicitly to the formalism of Ref.~\cite{abou1973selfconsistent}.}

\section{The integral equation for $\lambda(s)$ and the Anderson transition}
\label{sec:inteq}

The product of $G_i^{2s}$ along a path is a product of correlated random variables. {In the forward or upper limit approximation, these correlations are neglected and the propagators are treated as independent. }
To go beyond this approximation, we notice that two consecutive propagators along a path are related by the equation
\begin{equation}
 G_{k+1}=-\frac{1}{\epsilon + \zeta + G_k},
\end{equation}
where $\epsilon$ is the onsite energy at site $k$, and $\zeta= \sum_{j=1}^{K-1} G_j$ with $G_j$ i.i.d.\ random variables with distribution $P(G)$. The conditional probability
\begin{equation}\label{eq:KernelZero}
 P(G_{k+1} | G_k)= \mathbb{E}_{\epsilon, \zeta} \quadre{\delta \tonde{G_{k+1} + \frac{1}{\epsilon + \zeta + G_k}}}
\end{equation}
defines an integral operator $\hat{K}$ with the non-symmetric kernel $K(y,x)\equiv P(y|x).$ The operator is Markovian, since
\begin{equation}\label{eq:MarkovOp}
 \int_{-\infty}^\infty dy K(y,x)=1.
\end{equation}
It thus defines a continuous state Markov chain. Note that the chain is non-reversible, since no detailed balance condition holds; however, the following symmetry \cite{abou1973selfconsistent,altshuler1989distribution} holds:
\begin{equation}\label{eq:symm0}
 K(x,y)= \frac{1}{x^2 y^2} K \tonde{\frac{1}{y}, \frac{1}{x}}.
\end{equation}
The stable distribution $P(G)$ solving (\ref{eq:PGeq}) is a right eigenvector of $\hat{K}$ with eigenvalue $1$; the corresponding left eigenvector is the function with constant value $1$, as seen from Eq.\ \eref{eq:MarkovOp}.

The Kernel $K$ can be used to compute $\lambda(s)$ by writing
\begin{eqnarray}
\langle \chi^{s}\rangle=\langle G^{2s}_L...G^{2s}_1\rangle\nonumber=\int dG_{L-1} ... dG_1\ G_L^{2s} K(G_L,G_{L-1}) ...G_1^{2s}K(G_1,G_0)\ P(G_0).
\label{eq:kerker}
\end{eqnarray}
Introducing the vector space notation:
\begin{equation}
 \langle x| P \rangle= P(x), \quad \langle P | x \rangle =1, \quad \langle y| X| x \rangle= x \delta(x-y),
\end{equation}
we have
\begin{equation}
\langle \chi^{s} \rangle= \langle P |(K X^{2s})^L| P \rangle.
\end{equation}
Following the usual arguments on transfer matrices, we see that $\lambda(s)$ is the largest eigenvalue of the integral Kernel
\begin{equation}
K_s(y,x)\equiv K(y,x)x^{2s},
\label{eq:Ks}
\end{equation}
\emph{i.e.}, it is the largest solution of the equation
\begin{equation}
\lambda(s)\phi_s(y)=\int dx K_s(y,x)\phi_s(x).
\label{eq:Ks_int}
\end{equation}
This integral equation is exactly the same one that appears in Ref.~\cite{abou1973selfconsistent}, where it was derived by investigating the tails of the probability density of the imaginary part of the self energies (equivalently, the behavior of its characteristic function close to the origin).

The integral operator can also be exploited to determine the correlation functions between subsequent cavity propagators along a path in the lattice. Since $K$ is a contraction, being the average of contractions, then $\lambda=1$ is its largest eigenvalue. Let $\lambda_1$ be the second largest eigenvalue, with right and left eigenvectors $\phi_1, \psi_1$ respectively and assume there is a finite gap between the two:
\begin{equation}\label{eq:deco}
 K= |P \rangle \langle P | + \lambda_1 | \phi_1 \rangle \langle \psi_1 |+...
 \end{equation}
The connected part of the correlation function
\begin{equation}
  \langle G_r^{2s} G_0^{2s} \rangle= \langle P | X^{2s} K^r X^{2s} | P \rangle= \langle G_r^{2s}\rangle\langle G_0^{2s}\rangle+    \langle G_r^{2s} G_0^{2s} \rangle_c
\end{equation}
decays exponentially with $r$, as follows from Eq.\eref{eq:deco}, with leading behavior
\begin{equation}
   \langle G_r^{2s} G_0^{2s} \rangle_c = \lambda_1^r \langle P| X^{2s}|\phi_1 \rangle \langle \psi_1 | X^{2s}| P \rangle+ ...\ .
   \label{eq:ggc}
\end{equation}
that is dictated by the second largest eigenvalue of the kernel $K$, irrespectively of the value of $s$, assuming of course that the quantity on the right-hand side of (\ref{eq:ggc}) is finite (which sets the bounds $0\leq s<1/2$). One can define a correlation length $\xi=-1/\ln\lambda_1$, and show (analytically for the Cauchy-Lorentz disorder and numerically for box disorder) that it is always $\lambda_1<1$, which means that $\xi$ is always finite. In this respect the Anderson model is not different from other statistical models on the Bethe lattice in that the correlation length does not diverge.

\subsection{Box disorder distribution}
For box disorder in $[-W/2, W/2]$, we compute the kernel by first determining the probability distribution function $P(G)$ of the cavity propagators with population dynamics, as outlined in Sec. \ref{sec:PD}. We use a cubic spline interpolator, which is efficient to integrate numerically, plus the Cauchy-like tails $\sim G^{-2}$ for $P(G)$. We then compute \eref{eq:Ks} on a discretized $(x,y)$ grid:
\begin{equation}
K_s(x,y)=\int_{-W/2}^{W/2} d\epsilon \frac{\abs{x}^{2s}}{Wy^2}{\mathcal{P}_{\zeta}}\tonde{\frac{1}{y}-\epsilon-x},
\label{eq:K}
\end{equation}
{where  $\mathcal{P}_\zeta(\cdot)$ denotes the distribution of  $\zeta= \sum_{j=1}^{K-1} G_j$.}
The discretization is parametrized with $(x,y)=(\tan \theta_1, \tan \theta_2)$ where $\theta\in[-\pi/2, \pi/2]$ with an equal spacing in both directions $\Delta\theta=10^{-3}$.\footnote{We have changed $\Delta \theta$ over the range $[0.5\cdot 10^{-3},2\cdot 10^{-2}]$ and for population size $N=3\cdot 10^8$. The dependence on $\Delta \theta$ is quite small (like $\Delta \theta^3$) and the estimated error we associate to the discretization is of one part in $10^4$.}

We compute the largest eigenvalue $\lambda^{(N)}$ (for a population of given size $N$) of the discretized matrix $K_s$ by means of exact diagonalization, with the Arnoldi iterative algorithm. For $s=0$ and $1$, we know that $\lambda=1$; we use this as a criterion for the goodness of the discretization of $K$ (other errors, coming from the numerical integration of $P(G)$ and the exact diagonalization, are otherwise well controlled), resulting in an error less than $0.1\%$ for the chosen value of $\Delta \theta=10^{-3}$ and population size $N = n s = 10^9$. {The dependence of the resulting $\lambda^{(N)}$ on the population size $N$ (which led in a previous version of this manuscript to overestimate $W_c$ by about $1.5\% $) is significantly reduced if one recognizes that asymptotically $P(G)\propto a/G^2$ and that this asymptote is quickly reached for $G=O(1)$. The coefficient $a$ can then be extracted from a fit to typical values of $G$ with high accuracy. Values of $sN=10^9$ used here are then more than sufficient to reach the precision quoted in this paragraph.\footnote{If one neglects this information about the functional form of the tail of $P(G)$, and fits considering not typical but extreme values of $G$, a dependence of $\lambda^{(N)}=c_0+c/\log(N)^2$ is observed for $N<N^*$ which however saturates to $\lambda^{\infty}$ for $N\gg N^*$. This dependence can be traced back to the existence of a critical volume which scales like $e^{C/|W_c-W|^{1/2}}$ as $W\to W_c$.}}

\begin{figure}[htbp]
\begin{center}
\includegraphics[width=.495 \columnwidth]{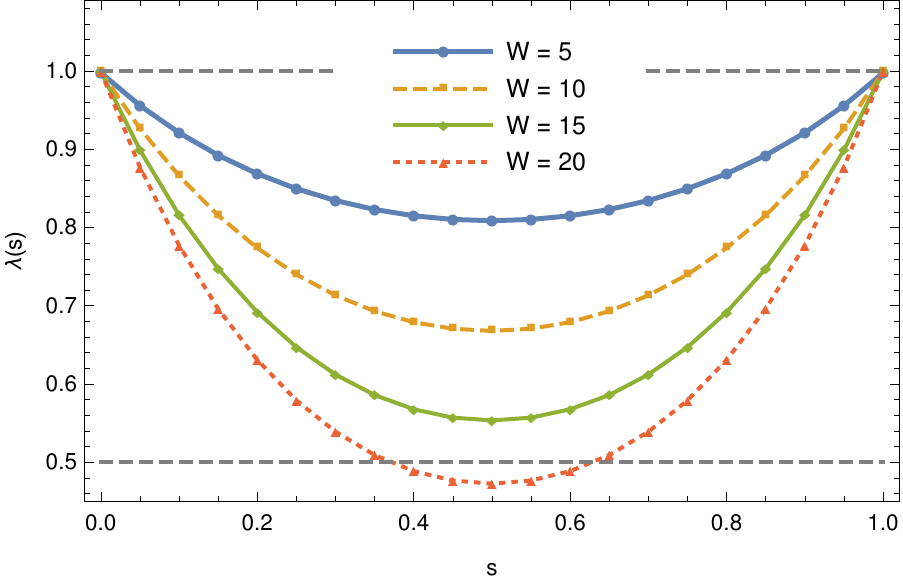}
\includegraphics[width=.495\columnwidth]{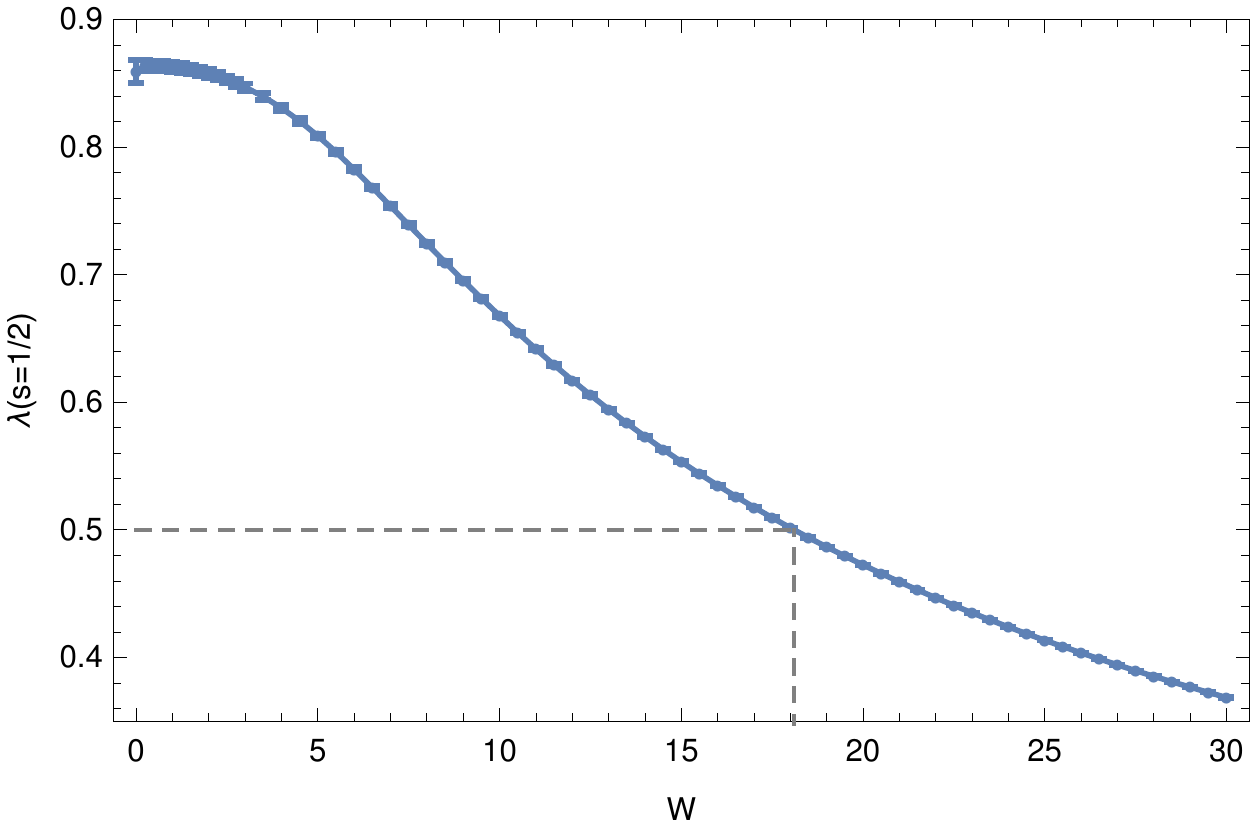}
\caption{{\it Left panel.} Largest eigenvalue $\lambda(s)$ as a function of the exponent $s$ of Eq. (\ref{eq:chis}), showing the symmetry at $s=1/2$. {\it Right panel.}  Largest eigevalue $\lambda(s)$ of the kernel $K_s$ for $s=1/2$, as a function of disorder $W$. The value of $W$ for which $\lambda(s=1/2)=1/2$ corresponds to the localization transition (here $W_c=18.11\pm0.02$).}\label{fig:lambdas}
\end{center}
\end{figure}

In Fig. \ref{fig:lambdas} we plot the resulting largest eigenvalue $\lambda(s)$ as a function of the exponent $s$. Since the minimum at $s=1/2$ is the value of interest for the transition, we also plot $\lambda(s=1/2)$ as a function of disorder. Here, the crossing of $\lambda=1/2$ happens at $W_c=18.11\pm0.02$, which is the localization transition point. {This is in agreement with the most recent numerical results in \cite{MirlinCriticalBethe}}.\\

We now consider the behavior of the typical value $\theta_{\rm typ}= \partial \lambda(s)/ \partial s |_{s=0}$ at small values of disorder. For box disorder $\partial \theta_{\rm typ}/\partial W=0$ at $W=0$, and we conjecture that this happens for all disorder distributions which fall to zero faster than $1/\epsilon^2$ at large $\epsilon$. We can show this by computing $\partial P(G)/\partial W$ at small $W$. The integral equation to be solved is
\begin{eqnarray}
&\frac{\partial P}{\partial W}(G) =  \int d^K G\int d\epsilon\frac{\partial\rho}{\partial W}(\epsilon)\delta\left(G+\frac{1}{\epsilon+\sum_{i=1}^K G_i}\right) P(G_1)...P(G_K)\\
 & + K\int d^K G\int d\epsilon\rho(\epsilon)\nonumber\delta\left(G+\frac{1}{\epsilon+\sum_{i=1}^K G_i}\right)  P(G_1)...P(G_{K-1})\nonumber\frac{\partial P}{\partial W}(G_K).
\end{eqnarray}
This is a linear integral equation (a tangent map) for $v(y)=\frac{\partial P}{\partial W}(y)$.
For $W=0$ we can use the result (\ref{eq:PGCclean}) and compute the integrals explicitly. The second integral can be performed easily (as $\rho(\epsilon,W=0)=\delta(\epsilon)$) while in the first one it is convenient to keep $\rho(\epsilon)=W^{-1}\theta(W/2-|\epsilon|)$ in the integral, and then expand for small $W$.

After some algebra one finds, for $K=2$ and $W\ll 1$:
\begin{eqnarray}
v(y)&=&W\frac{y^2 \left(3-2 y^2\right)}{3 \sqrt{2} \pi  \left(2 y^2+1\right)^3}+
\int dx\frac{\sqrt{2}}{\pi  \left(2 x^2 y^2+4 x y+y^2+2\right)}v(x).
\label{eq:intvx}
\end{eqnarray}
We have
\begin{eqnarray}
\frac{\partial \theta_{\rm typ}}{\partial W}=-\int dx (\ln x^2) v(x).
\end{eqnarray}
Without solving the equation (\ref{eq:intvx}) one can see that $v(x)=O(W)$ so we have $\theta_{\rm typ}(W)\propto W^2$ for small $W$ (up to constants that do not depend on $W$). Notice that this is only true for box disorder: for Cauchy disorder the corresponding first term in the right-hand side of (\ref{eq:intvx}) is $O(1)$ instead of $O(W)$ and this gives $\theta_{\rm typ}(W)\propto W$, see the following section.

\subsection{Cauchy on-site randomness}
In the case of Cauchy on-site randomness,
\begin{equation}
\rho(\epsilon)=\frac{W/\pi}{W^2+\epsilon^2}.
\label{eq:rhoCauchy}
\end{equation}
the integral operator with Kernel \eref{eq:KernelZero} (corresponding to $s=0$) can be explicitly diagonalized.
In this case the equilibrium distribution of the cavity fields at $E=0$ is itself a Cauchy distribution
\begin{equation}
    P(G)= \frac{1}{\pi} \frac{K \alpha + W}{(K \alpha + W)^2 G^2 + 1},
\end{equation}
with width
\begin{equation}\label{eq:alpha}
 \alpha= \frac{W}{2 K}\tonde{-1+ \sqrt{1+ \frac{4 K}{W^2}}}.
\end{equation}
Plugging this into the definition of the Kernel one finds
\begin{equation}
 K(y,x)= \frac{1}{\pi} \frac{\overline{\alpha}}{\overline{\alpha}^2 y^2 + \tonde{1 + xy}^2},
\label{eq:Kcauchy}
\end{equation}
where the two constants $\alpha, \overline{\alpha}$ are related by:
\begin{equation}\label{eq:alpharel}
 \overline{\alpha}=\frac{1-\alpha^2}{\alpha}, \quad  \alpha=- \frac{\overline{\alpha}}{2} + \frac{1}{2}\sqrt{{\overline{\alpha}}^2+4}.
\end{equation}
The Kernel, being not symmetric, has distinct left and right eigenvectors. As we show in the Appendix, the eigenvalues are:
\begin{equation}
\lambda_n=(-1)^{n}\alpha^{2n}
\label{eq:lambdaCs0}
\end{equation}
for $n=0, 1, \cdots$, where  $\lambda_0 \equiv \lambda=1$. The connected component of the correlation function between the square modulus of the cavity Green functions along a path decays over a correlation length that is related to the spectral gap and reads
\begin{equation}
\xi=-\frac{1}{2\ln\tonde{\frac{W}{2K}\tonde{\sqrt{1+\frac{4K}{W}}-1}}}.
\label{eq:xi}
\end{equation}
In the clean case $W=0$, we recover $\alpha=1/\sqrt{K}$.

For $n=1,2, \cdots$ the right and left eigenfunctions ($\Phi^{(n)}(x)$ and $\Psi^{(n)}(y)$, respectively) can be compactly written as follows:
\begin{eqnarray}\label{eq:RightEigen}
\Phi^{(n)}(x)=&C_n \sum_{m=0}^{n-1} \quadre{\frac{-\alpha^2}{1+ \alpha^2}}^{n-1-m}{n \choose m+1} \nonumber\\
&\frac{1}{m!}\lim_{t \to 0} \frac{d^m}{dt^m} \quadre{\frac{-\alpha^2(1-t)^2 x^2+ (\alpha^2+t(1-\alpha^2))^2}{\quadre{\alpha^2(1-t)^2 x^2+ (\alpha^2+t(1-\alpha^2))^2}^2}},
\end{eqnarray}
and
\begin{eqnarray}\label{eq:LeftEigen}
    \Psi^{(n)}(y)= &- \frac{1}{(1+ \alpha^2)^n}+\sum_{m=0}^{n-1} \tonde{\frac{-\alpha^2}{1+ \alpha^2}}^{n-1-m} {n \choose m+1}   \nonumber\\
   & \frac{1}{m!}\lim_{t \to 0} \frac{d^m}{dt^m} \quadre{\frac{1}{1-t} \frac{1}{1+\alpha^2 y^2 (1-t)^2}},
\end{eqnarray}
as we show in the Appendix. In \eref{eq:RightEigen}, the normalization constant $C_n$ is chosen in such a way that $\int_{-\infty}^{\infty} dx \Phi^{(n)}(x) \Psi^{(n)}(x)=1$.

The typical value of the path susceptibility is readily obtained as:
\begin{eqnarray}
\theta_{\rm typ}&=&-\bra{P}K\ln X^2\ket{P}
=-\int_{-\infty}^\infty dx\ln(x^2)\frac{\alpha/\pi}{\alpha^2+x^2}=-2\ln\left(\alpha\right),
\label{eq:theta_typ_cauchy}
\end{eqnarray}
which agrees with the expression in \cite{aizenman2011resonant} relating the typical value in presence and in absence of disorder in the case of a Cauchy distribution of the disorder.
Notice that since $\alpha\leq 1/\sqrt{K}$ (the equality being valid only at zero disorder), at $E=0$, $\theta_{\rm typ}>\ln K$ in the deep delocalized phase. In particular, for Cauchy disorder $\theta_{\rm typ}$ grows linearly with $W$,
\begin{eqnarray}
\theta_{\rm typ}=\ln K+\frac{W}{2\sqrt{K}}+O(W^3).
\label{eq:D1_cauchy}
\end{eqnarray}
We will discuss in the following section how, if  $\theta_{\rm typ}$ is connected to the fractal dimension $D_1$ of the wave function of a finite tree-like lattice as suggested in Ref. \cite{kravtsov2017non}, this would imply that the wave function is \emph{never} ergodic.

\subsection{On the fractal dimensions of non-ergodic extended states}
\label{sec:computationoffractal}
In several recent works \cite{Biroli:2012vk, de2014anderson, kravtsov2017non} it has been suggested that the delocalized phase of the Anderson model on the RRG with $N$ sites has two distinct phases: one in which the participation ratios of the eigenfunctions is $\propto N$ and one in which it is $\propto N^\alpha$ for some $\alpha<1$. This was questioned in several other papers \cite{tikhonov2016anderson,garcia2017scaling, TikhMirlin2018} (however a genuine multi-fractal region was shown to exist for the finite Cayley tree with $N$ vertices \cite{PhysRevB.94.184203, SonnerTikhonov,  MonthusMultifractal}, and to control the RRG properties below a given critical volume \cite{BiroliTarzia2018}); to the best of our understanding the question is whether this phase exists on the RRG or it is washed away when going from the Cayley tree to the RRG.

This question can not be addressed within the framework discussed in this work. However, we can point out the connection between some arguments presented in Ref.~\cite{kravtsov2017non} and the quantities investigated here, in particular the largest eigenvalues $\lambda(s)$ of the integral equation (\ref{eq:Ks_int}). We will do it assuming the definition of the fractal dimension $D_1$ given in Eq.\ (76) of Ref. \cite{kravtsov2017non}:
\begin{eqnarray}\label{eq:KrD1}
D_1=\frac{\Lambda}{\ln K}
\end{eqnarray}
with
\begin{eqnarray}
\Lambda&=&\min_{s\in(0,1]}\lim_{\ell\to\infty}\frac{1}{s\ell}\ln(K^\ell\langle\prod_{i=1}^\ell |G_i|^{2s}\rangle)= \min_{s\in (0,1]}\frac{\ln K+\mu(s)}{s}
\end{eqnarray}
where $\mu(s)=\ln\lambda(s)$ has already appeared in Eq.~(\ref{eq:Qchi}). The extremization procedure with respect to $s$ follows from the interpretation of $\Lambda$ as a free energy of a directed polymer model~\cite{Biroli:2012vk}, which undergoes a 1-step RSB transition. The cavity propagators $G_i$ in this equation are \emph{not} independent, but follow the recursion relation (\ref{eq:Gcav}), and one can find $\lambda(s)$ by solving the integral equation for $K_s$.

For $W=0$, the minimum is attained at $s^*=1$; since $\mu(1)=\mu(0)=0$ (for any $W$) this gives $D_1=1$, which is interpreted as a condition for a fully ergodic state. In \cite{kravtsov2017non} it is shown that, when using a slightly modified version of the large disorder forward scattering approximation\footnote{The conventional forward scattering approximation consists in neglecting the term $\sum G_k(E)$ (which are self-energy corrections) in the denominator of the recursive equation for the cavity Green functions, thus decoupling the consecutive propagators along the path and leading to {$G_i(E) \sim (E- \epsilon_i)^{-1}$}. In \cite{kravtsov2017non} this approximation is performed, but the distribution of the on-site randomness $\epsilon_i$ is modified introducing a cut-off around the origin, in order to partially account for the regularization effect produced by the self energies (or, equivalently, of for the anti-correlations between consecutive propagators along the path), in the same spirit of the \emph{lower limit} approximation in \cite{Anderson1958}}, the same remains true for sufficiently small $W\leq W_{erg}\simeq 5.74$ (for $K=2$), meaning that the minimum is still achieved at $s^*=1$ for all these values of disorder (and thus $D_1=1$).

As $W$ is increased the minimum moves away to $s=s^*<1$, and we have
\begin{eqnarray}
0=\frac{\partial}{\partial s}\frac{\ln K+\mu(s)}{s}=\frac{-\ln K-\mu(s)+s\mu'(s)}{s^2}
\end{eqnarray}
which gives
\begin{eqnarray}
\Lambda=\mu'(s^*).
\end{eqnarray}

\begin{figure}[htbp]
\begin{center}
\includegraphics[width=.5\columnwidth]{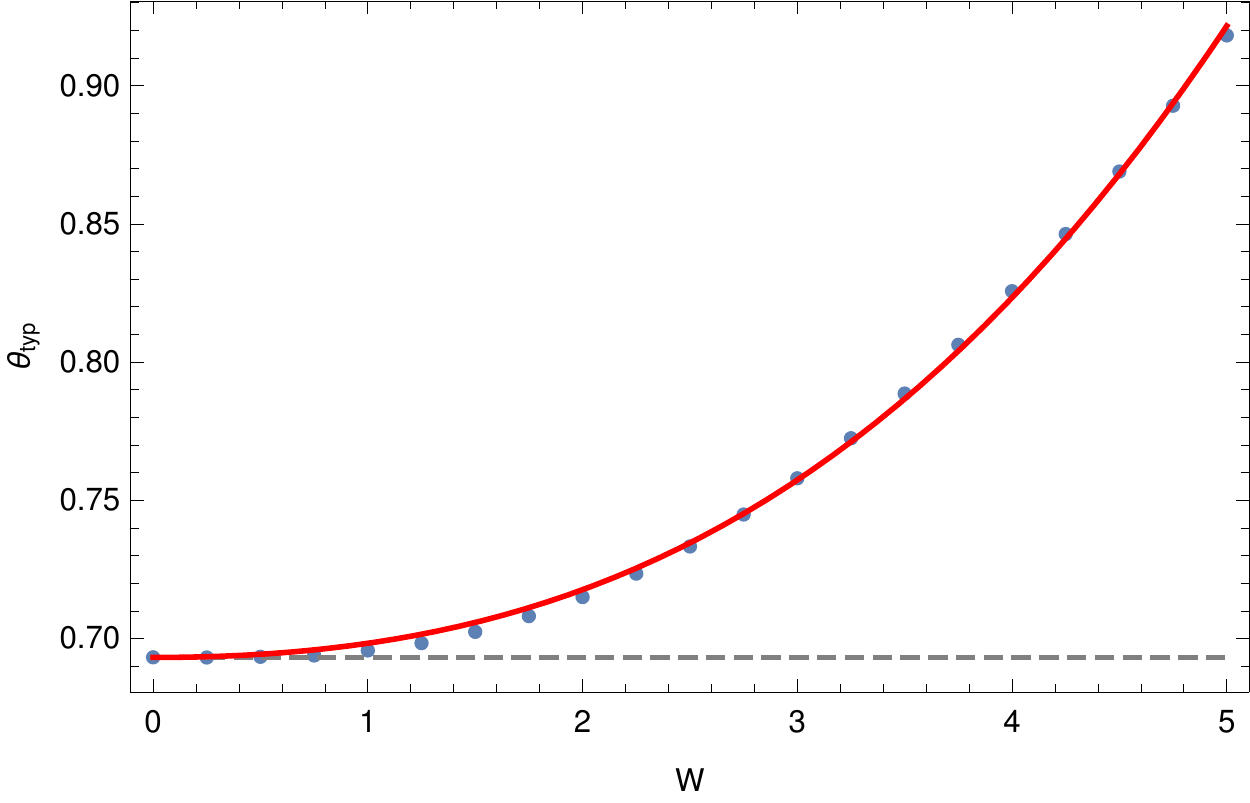}
\caption{$\theta_{\rm typ}$ as a function of disorder for a box disorder distribution. The red line is a fit of the form $\theta_{\rm typ}=\log 2 + c W^2 + d W^3$, with $c=(4.1 \pm 0.2) \cdot 10^{-3}$ and $d=(1.00 \pm 0.06) \cdot 10^{-3}$.}
\label{fig:muprime}
\end{center}
\end{figure}

Notice that, due to the symmetry $\mu(s)=\mu(1-s)$ we have, for $s^*\simeq 1$ and $W\gtrsim W_{erg}$,
\begin{eqnarray}
\Lambda\simeq -\mu'(0)=\theta_{\rm typ}.
\end{eqnarray}
Now, one can argue that $W_{erg}>0$ only if $\theta_{\rm typ}=\ln K$ at $W=W_{erg}$. This is indeed what happens within the approximation exploited in \cite{kravtsov2017non}. However, this result is a feature of that approximation, as we illustrate here. We first notice that, for the Cauchy distribution, $\theta_{\rm typ}=\ln K$ \emph{only} at $W=0$ (see Eq.(\ref{eq:theta_typ_cauchy})) and that for $W>0$ the strict inequality $\theta_{\rm typ}>\ln K$ holds. Second, we numerically check, by using the equality $\theta_{\rm typ}=-\langle\ln G^2\rangle$, that the same holds also for the box distribution, although $\partial \theta_{\rm typ}/\partial W=0$ at $W=0$ for box disorder (see Fig. \ref{fig:muprime}). From this we are led to conjecture that (provided that Eq. (\ref{eq:KrD1}) is used to define the fractal dimension),
$
D_1<1
$
for all $W>0$ and for all disorder distributions. We also conjecture that for all disorder distributions which fall faster than $1/\epsilon^2$ at large $\epsilon$, $\theta_{\rm typ}=\ln K+c W^2$.

Under these hypothesis we now show that
\begin{eqnarray}
D_1=1- \kappa W^4+o(W^4) ,
\end{eqnarray}
as numerically found in Ref.\cite{kravtsov2017non} (for box disorder we find numerically $\kappa=(5.0 \pm 0.6) \cdot 10^{-5}$). In fact using $\lambda(s)=\lambda(1-s)$ one can write, close to $s=1$, and by defining $s=1-\sigma$,
\begin{eqnarray}
\Lambda&=&\min_{\sigma\in[0,1)}\frac{1}{1-\sigma}(\ln K+\mu(\sigma))\\
&=&\min_{\sigma\in[0,1)}\frac{1}{1-\sigma}(\ln K+\mu(0)+\sigma \mu'(0)+
\frac{1}{2}\sigma^2\mu''(0)+...),
\end{eqnarray}
and considering that $\mu(0,W)=0$, $\mu'(0,W)\equiv \frac{\partial \mu(\sigma,W)}{\partial \sigma}|_{\sigma\to 0}\simeq -\ln K -c W^2$, and $\mu''(0,W)>0$ (numerically we find $c=(4.1 \pm 0.2) \cdot 10^{-3}$), for small $W$ one has
\begin{eqnarray}
\Lambda=\ln K-\frac{c^2 W^4}{2\mu''(0,0)},
\end{eqnarray}
or
\begin{eqnarray}
D_1=1-\frac{c^2}{2\mu''(0,0)\ln K} W^4+o(W^4).
\end{eqnarray}
Notice that the constant $c$ can be computed using perturbation theory in $s$ and $W$ around the $W=0,s=0$ point.
As the problem for $W=0$ is reduced to the Cauchy distribution, this can be done using the eigenvectors and eigenvalues of the kernel reported in the Appendix \cite{WorkInProgress}.

In conclusion, we believe that the quantity $D_1$, defined in \cite{kravtsov2017non} by using quantities computed on the tree, is always analytic, with only 3 vanishing derivatives at $W=0$. One is led to conclude that the full inclusion of the self-energy corrections in the computation of $\lambda(s)$ breaks the analogy with directed polymers with independent on-site energies, so that the RSB phase present in that directed polymer calculation is not present in the Anderson model at any finite $K$. The fact that $D_1$ is very flat at the origin can be considered a vestige of that phase and it is somewhat surprising that a simple computation in the approximation of \cite{kravtsov2017non} captures the qualitative feature of a very flat $D_1$ at $W=0$.

\section{The large disorder, large connectivity limit: an interpolation formula}
\label{sec:largedis}

In this section we give a {concise} proof of the asymptotic formula $W_c=4K \ln K$ for the Anderson transition point $W_c$  (this result is specialized for box disorder, the general formula is given in Eq. (\ref{eq:transition})).
This{ formula is known}  since the original work of Anderson \cite{Anderson1958} and has been rigorously proved in Ref.~\cite{Bapst2014HighConnectivity}, to which we refer for a more comprehensive list of references. Additionally, we provide an interpolating formula between the exact result and the ``upper bound'' formula $W_c=2 e K\ln K$ \cite{Anderson1958} obtained within the forward approximation (FA) \cite{de2014anderson, pc2016forward, kravtsov2017non}.  Due to the central importance of the FA for analytic treatments of many-body localization \cite{Basko:2006hh, ros2015integrals, SmallBath, imbrie2017review} we believe it is important to understand how one could go beyond it in a controlled expansion at large disorder/large connectivity.

To begin with, we notice that the distribution $P(G)$ changes continuously with $W$ for any $K$: what signals the transition is the divergence with $L$ of the \emph{typical} value of the susceptibility $\chi$ summed over $K^L$ paths. Therefore, one can make some assumption on $P(G)$ and use it in the kernel $K_s$. In the limit of large connectivity, $K\gg 1$, if we are interested in the transition we have $W=O(K)$ and therefore some simplifications occur. One can self consistently neglect $\sum_{i=1}^K G_{i}$ with respect to $\epsilon_i$, since the former turns out to be of $O(K/W)$, while the latter is of $O(W)$. This is licit as long as $W\gg \sqrt{K}$. Notice that this was the scale of disorder on which $D_1$ decayed from its maximum value $D_1=1$ at $W=0$. Numerically, this is also not far from the disorder value at which the maximum of $P(G)$ moves away from $G=0$.
Under these hypotheses, the solution for $P(G)$ is
\begin{equation}
P(G) = \int d\epsilon \rho(\epsilon)\delta\left(G-\frac{1}{\epsilon}\right)=\frac{1}{G^2}\rho \left(\frac{1}{G} \right).
\label{eq:applargW}
\end{equation}
In the case of Cauchy distribution this functional form is exact and this approximation amounts to replacing $\alpha\to 1/W$:
\begin{equation}
P(G)=\frac{1}{G^2}\frac{W/\pi}{W^2+1/G^2}=\frac{(W^{-1})/\pi}{W^{-2}+G^2}.
\label{eq:pg}
\end{equation}
In the case of box distribution, Eq.\ (\ref{eq:applargW}) turns out to be a very good approximation already for $K=2, W\gtrsim 10$.

Let us now show that, for $W\gg 1$, $P(G)$ is also \emph{the only eigenvector corresponding to a non-zero eigenvalue} of the kernel $K_{1/2}$. {Following the same reasoning as above, we can neglect the factor $\zeta= \sum_{j=1}^{K-1}G_j$ in the definition of the Kernel: this corresponds to replacing $\mathcal{P}_\zeta$ in Eq. (\ref{eq:K}) with a delta function in zero, so that the Kernel simplifies to:
\begin{equation}
K_{s=1/2}(y,x)=\int d\epsilon \, \rho(\epsilon) |x| \delta \tonde{y+ \frac{1}{x+ \epsilon}}.
\end{equation}}
We then need to solve the integral equation
\begin{eqnarray}
\lambda(1/2)\phi_{1/2}(y)&=&\int d\epsilon \rho(\epsilon)\int dx\ \delta\left(y+\frac{1}{x+\epsilon}\right)|x|\phi_{1/2}(x)\nonumber\\
&=&\int dx\ \frac{1}{y^2}\rho\left(x+\frac{1}{y}\right)|x|\phi_{1/2}(x),
\label{eq:lambdaphi}
\end{eqnarray}
where in the last line we have integrated with respect to $\epsilon$.

In order to prove that $P(G)$ is an eigenvector of this integral equation, after factoring out $1/y^2,$ we need to show that $\mathcal{I}\propto\rho(1/y)$ where
\begin{equation}
\mathcal{I}=\int dx \rho\left(\frac{1}{y}+x\right)|x|\frac{1}{x^2}\rho\left(\frac{1}{x}\right),
\label{eq:i}
\end{equation}
and the proportionality coefficient is $\lambda(1/2)$. As a first step, it is convenient to pass to the Fourier transform, by defining
\begin{equation}
\rho(x)=\int\frac{dk}{2\pi}\tilde\rho(k)e^{ikx}
\label{eq:rho}
\end{equation}
to get
\begin{equation}
\mathcal{I}=\int \frac{dk}{2\pi}e^{ik/y}\tilde\rho(k)\int\frac{dk'}{2\pi}\tilde\rho(k')\int_{-\infty}^\infty dx e^{ikx+ik'/x}\frac{1}{|x|}
\label{eq:i_2}
\end{equation}
The integral can be done in terms of Bessel functions:
\begin{eqnarray}
&&\int_{-\infty}^\infty dx e^{ikx+ik'/x}\frac{1}{|x|}=2\int_{0}^\infty \frac{dx}{x} \cos(kx+k'/x)\nonumber\\
&=&-4\frac{\partial J_n(2\sqrt{kk'})}{\partial n}|_{n\to 0}=-2\pi Y_0(2\sqrt{kk'}),
\label{eq:bessel}
\end{eqnarray}
where for $kk'<0$ we must take only the real part of the Bessel function $Y_0$ of complex argument.

If the disorder $W$ is the only scale of the distribution,  appearing in the functional form
\begin{equation}
\rho(x,W)=\frac{1}{W}\rho(x/W),
\label{eq:rho_2}
\end{equation}
then we have
\begin{equation}
\tilde\rho(k,W)=\tilde\rho(kW).
\label{eq:tilderho}
\end{equation}
Assuming that $\tilde\rho$ decays naturally on scales of $O(1)$, for large $W$ the integrals over $k,k'$ are cut-off at $1/W\ll 1$. We can then expand the Bessel function to find the leading divergence at small $k,k'$, which is logarithmic
\begin{equation}
\mathcal{I}=\int \frac{dk}{2\pi}e^{ik/y}\tilde\rho(k)\int\frac{dk'}{2\pi}\tilde\rho(k')(-2\log(|k k'|)+O(1)).
\label{eq:i_3}
\end{equation}
The integral over $k, k'$ is done by changing variables to $\theta=kW,\theta'=k'W$, keeping $Wy=O(1)$, while sending $W\to\infty$, considering that $\int \frac{d\theta}{2\pi W}e^{i\theta/Wy}\tilde\rho(\theta)=\rho(1/y)$ and $\int \frac{d\theta}{2\pi W}\tilde\rho(\theta)=\rho(0)$:
\begin{eqnarray}
\mathcal{I}&=&\int\frac{d\theta}{2\pi W}\frac{d\theta'}{2\pi W}e^{i\theta/(Wy)}\tilde\rho(\theta)\tilde\rho(\theta')(-2\log(|\theta\theta'|/W^2))\nonumber\\
&=&4\log(W)\rho(0)\rho(1/y)+O(1).
\label{eq:i_4}
\end{eqnarray}
So, to leading order in $\log(W)$, the eigenvalue is $\lambda(1/2)=4\log(W_c)\rho(0)$ and the transition point is located at
\begin{equation}
4\log(W_c)\rho(0)=\frac{1}{K}.
\label{eq:transition}
\end{equation}
For the box distribution this gives the familiar expression
\begin{equation}
\frac{4\log(W_c)}{W_c}=\frac{1}{K},
\label{eq:transitionbox}
\end{equation}
which is sometimes written in the form (equivalent to leading $O(\log(W))$)
\begin{equation}
W_c=4K\log(K).
\end{equation}

We can now pause and discuss the physical meaning of the fact that $\phi_{1/2}(x)=P(x)$ for large disorder. At the transition, the eigenvector $\phi_{1/2}$ is the conditioned probability of the propagator $G$, given that one of the $K^L$ paths gives a resonance \footnote{An alternative way to interpret this comes from realizing that the eigenvector $\phi_{1/2}(x)$ is proportional to $P(x)$, and the proportionality factor is the expectation value of the paths susceptibility $\chi_p$, \emph{conditioned} to the fact that the first cavity propagator along the path has amplitude $x$. A similar observation for the adjoint kernel is given in \cite{Bapst2014HighConnectivity}. The limit $\phi_{1/2}(x) \to P(x)$ translates into the fact that at the transition the expectation is dominated by instances where the susceptibility is $O(1)$ due to resonances, and for these events correlations between distant propagators do not matter (thus the conditioning is immaterial and its dependence on $x$ drops).}. The fact that $\phi_{1/2}=P$, which is the same distribution obtained by the forward approximation, means that one does not have to go to rare events to get a resonance at large $W$. Why then the FA does not give the correct limit for the transition point?

The reason is that the kernel corresponding to the FA, and that of $K_{1/2}$ share the eigenvector corresponding to the largest eigenvalue (in the large $W,K$ limit), but the eigenvalue itself is different (as well as the other eigenvectors and eigenvalues).

\begin{figure}[htbp]
\begin{center}
\includegraphics[width=.495 \columnwidth]{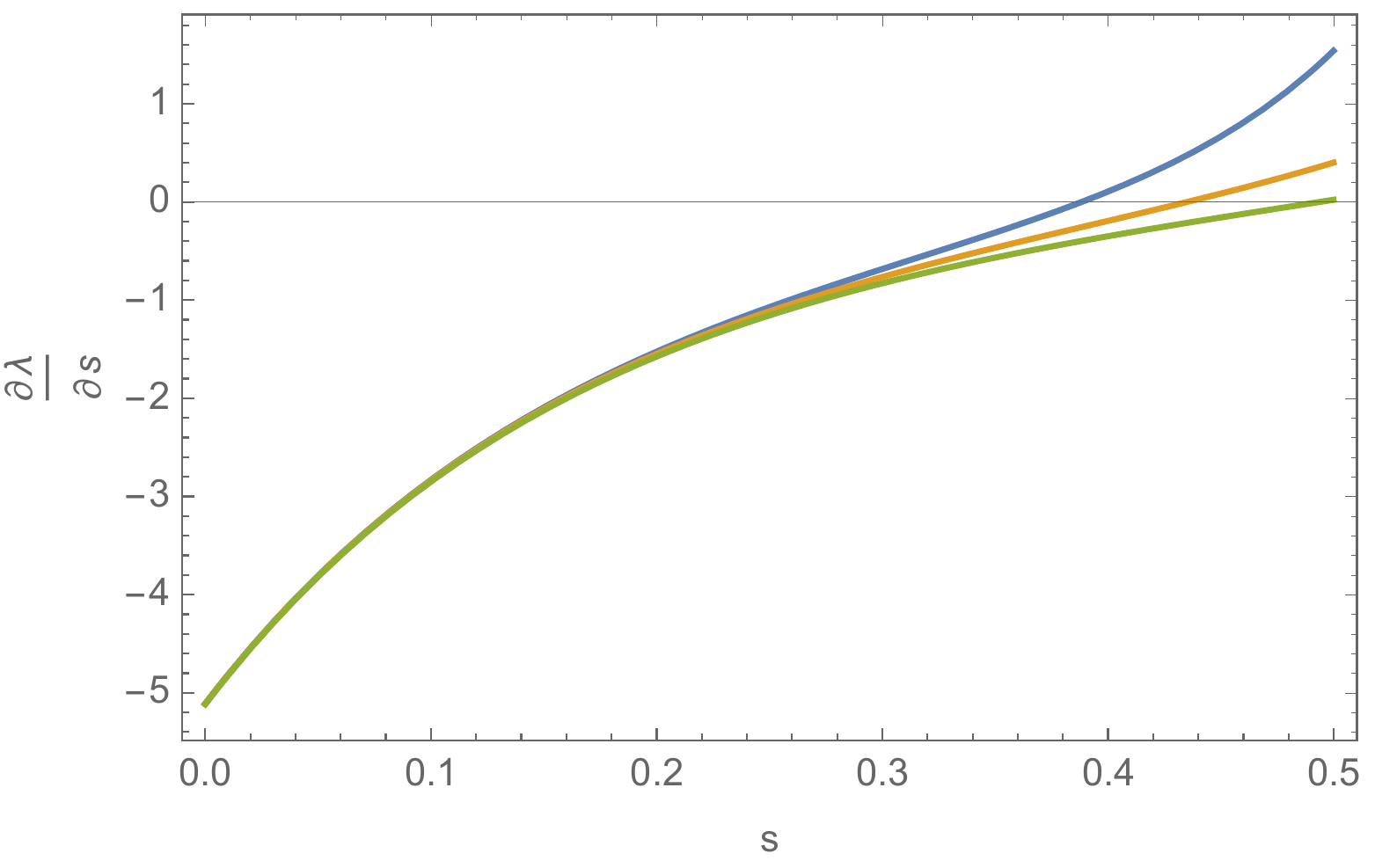}
\includegraphics[width=.495\columnwidth]{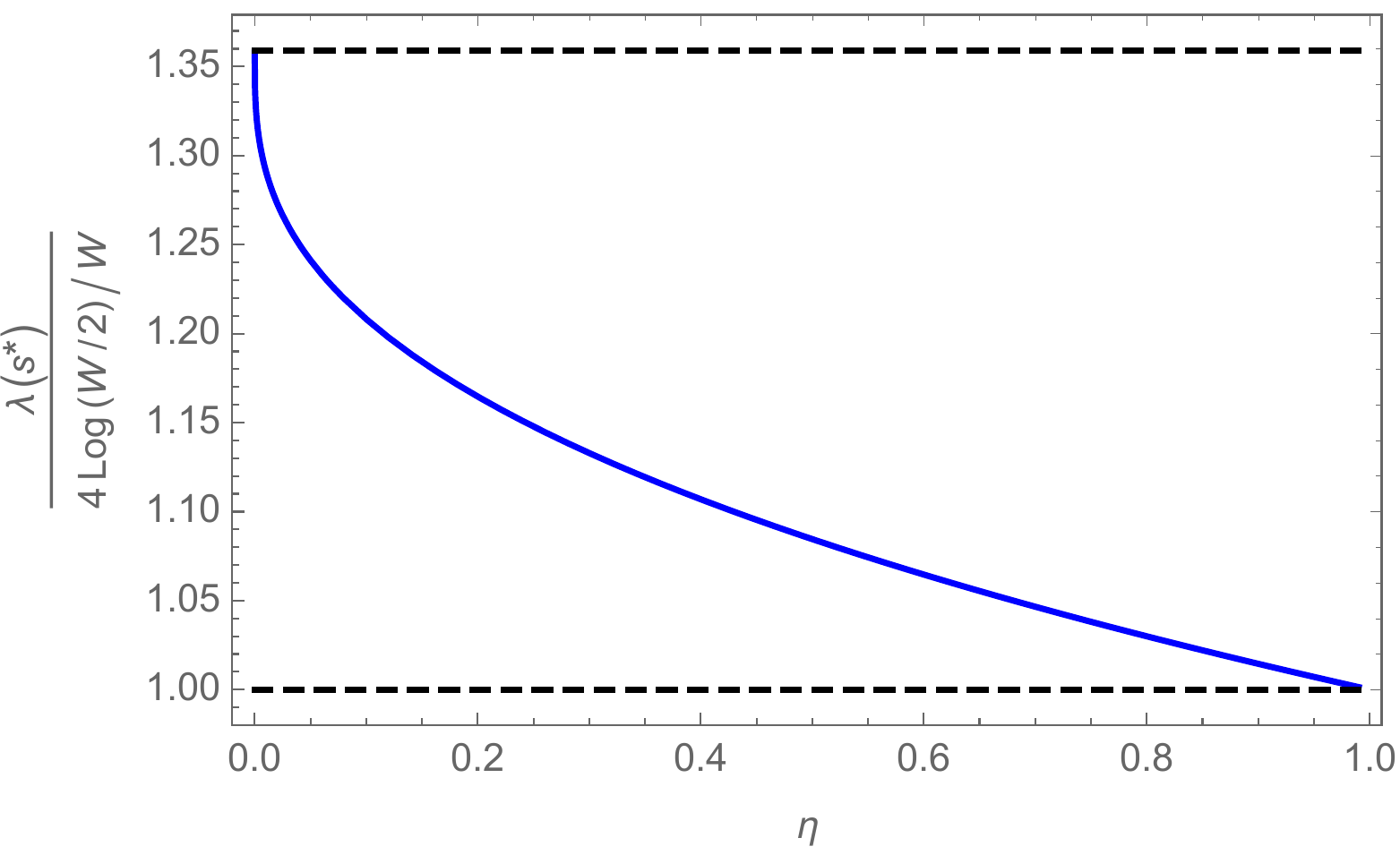}
\caption{{\it Left panel.} The point $s^*$ where $\lambda'(s^*)=0$ moves towards $s=1/2$ as $\eta=0.01, 0.2,0.99$ (blue, yellow, green respectively).  {\it Right panel.} The stationary value $\lambda(s^*)$ divided by $4\log(W/2)/W$ moves quickly from $e/2 \simeq 1.359$ to $1$ as $\eta\to 1$. We set $W=70$ for illustrative purposes.}
\label{fig:lambda-eta}
\end{center}
\end{figure}

Let us clarify this point by introducing the modified kernel
\begin{eqnarray}
K_{s}^\eta(y,x)&=&\int d\epsilon \rho(\epsilon) \delta\left(y+\frac{1}{\eta x+\epsilon}\right)x^{2s}{=}\frac{1}{y^2}\rho\left(\eta x+\frac{1}{y}\right)x^{2s},
\label{eq:Ks_2}
\end{eqnarray}
where $\eta\in[0,1]$. We have {$K^{\eta\to 0}=K^{\textrm{FA}}$}, corresponding to the forward approximation, since we are not considering the self-energy correction in the denominator. By contrast, $K^{\eta\to 1}=K$ is the full problem we have considered in this article. For any $\eta\neq 1$ the symmetry $\lambda(s)=\lambda(1-s)$ is not respected. So the transition point in the saddle point approximation
\begin{equation}
0=\theta=\mu'(s^*)
\label{eq:theta}
\end{equation}
is not achieved at $s^*=1/2$, but rather at a given point depending on $W$, see Fig.~\ref{fig:lambda-eta}. The simplest case, $\eta=0$ yields
\begin{equation}
K^0_{s}(y,x)=\frac{1}{y^2}\rho(1/y)x^{2s},
\label{eq:K0}
\end{equation}
which has again the eigenvector $\phi^0_{s}(x)=\frac{1}{x^2}\rho(1/x)$. The eigenvalue is
\begin{equation}
\lambda(s)=\int dx \rho(1/x)x^{2s-2}=\int d\epsilon\ \epsilon^{-2s}\rho(\epsilon).
\label{eq:lambdasint}
\end{equation}
Using the form $\rho(\epsilon)=\frac{1}{W}\rho(\epsilon/W)$ we have
\begin{eqnarray}
\lambda(s)=W^{-2s}\int dx x^{-2s}\rho(x)=W^{-2s}\langle x^{-2s}\rangle.
\end{eqnarray}
We need to minimize this expression with respect to $s$, and this yields the same result of the forward scattering approximation $\lambda(s^*)=2e\log(W/2)/W$, which for large $K$ gives the familiar
\begin{eqnarray}
W_c=2eK\ln K.
\end{eqnarray}

In the general case $\eta>0$, we should in principle solve the integral equation. However we can make a simplifying assumption: since $\frac{1}{x^2}\rho(1/x)$ is the eigenvector both for $\eta=0$ and $\eta=1$, one can assume that it is the eigenvector for any $\eta\in [0,1]$. In this way the eigenvalue is obtained by:
\begin{equation}
\lambda^\eta(s)=\frac{1}{\rho(0)}\int dx \rho(\eta x)x^{2s-2}\rho(1/x).
\label{eq:lambdaeta}
\end{equation}
For box disorder this reads
\begin{eqnarray}
\lambda^\eta(s)&=&\frac{2}{W}\int_{2/W}^{W/(2\eta)}dx\ x^{2s-2} {=}\frac{2^{-2 s} \left(16^s W^{2-2 s}-4 \eta  \left(\frac{W}{\eta }\right)^{2 s}\right)}{W^2 (1-2 s)}.
\label{eq:lambdaeta_2}
\end{eqnarray}
Notice that the symmetry $\lambda(s)=\lambda(1-s)$ holds \emph{only} for $\eta=1$.

Notice that, while for $\eta=0$ this yields an \emph{overestimate} of $W_c$, for $\eta=1$ it yields an \emph{underestimate}, which however becomes exact when $W\to\infty$. One can ask then what is the value of $\eta$ corresponding to the numerically obtained value for the smallest $K=2$. From our numerical investigations $W_c(K=2)=18.11$, which corresponds to {$\eta=0.836$}, quite close to $\eta=1$.

As a final result, notice that, for box distribution and $\eta=1$,
\begin{equation}
\lambda(s,W)=\frac{4\log(W/2)}{W}\frac{\sinh(2\sigma)}{2\sigma}
\label{eq:lambdasw}
\end{equation}
where
\begin{equation}
{\sigma=(s-1/2)\log(W/2)}
\label{eq:sigma}
\end{equation}
From this, we can recover the full $P(\chi)$ for any $W$. We quote here just the result for $\chi=O(1)$ and $W=W_c$, where we find the distribution of the single path susceptibility $\chi$,
\begin{equation}
P(\chi)=\frac{K^{-L}}{\chi^{3/2}}e^{-\frac{(\ln \chi)^2}{8L/(3(\ln (W/2))^2)}},
\label{eq:pchi}
\end{equation}
which is a log-normal distribution, in agreement with the general discussion {in Sec. \ref{sec:CritSusc}}. The typical values of this log-normal distribution are of order $\ln\chi\sim \sqrt{L}/\ln(W/2)$.

In the general case
\begin{equation}
\phi(s)=\ln\left(\frac{4\ln(W/2)}{4}\right)+\ln\left(\frac{\sinh(2\sigma)}{2\sigma}\right),
\label{eq:phis}
\end{equation}
and therefore the integral representation is
\begin{eqnarray}
P(\chi)=\ln(W/2)&&\left(\frac{4\ln(W/2)}{W}\right)^L\int_{B}\frac{d\sigma}{2\pi i}e^{L(\ln(\sinh(2\sigma)/2\sigma)-\sigma\tilde\theta)},
\label{eq:pchi_2}
\end{eqnarray}
where $\tilde\theta=\ln(W/2)\ln\chi/L$. A real saddle point exists only for $-2<\tilde\theta< 2$, so the main range of variability\footnote{For $\tilde\theta$ outside this range there is a pair of complex conjugate saddle points and the resulting integral is much smaller. So we can speak of ``main range'' for $\chi$.} of the susceptibility is $\chi\in [e^{-\frac{2 L}{\ln(W/2)}},e^{+\frac{2 L}{\ln(W/2)}}].$

\section{{Summary and perspectives}}
\label{sec:conclusions}
In this article we have revisited the Anderson transition on the Bethe lattice in terms of real-energy propagators. We have first determined the distribution of these propagators through population dynamics and compared it with the one resulting from exact numerical methods on Random regular Graphs, finding agreement.
We have formulated a novel criterion for the transition in terms of the distribution of the susceptibilities, writing an integral equation for the quantity $\lambda(s)$
determining the Mellin transform of the distribution; we have shown that this criterion is consistent with the previously known one, given that this equation coincides with the one originally introduced in \cite{abou1973selfconsistent} to determine the stability of the Anderson phase.

In the case of Cauchy disorder, we have determined analytically the full set of eigenvalues and eigenvectors of the integral kernel at $s=0$, and we have shown explicitly that the gap between the largest eigenvalues of the integral kernel is always $>0$, implying that the cavity propagators have exponentially decaying correlations. For box disorder we have derived an accurate estimate of the transition point for lattices with connectivity $k=2$, and we have shown analytically that the fractal dimension $D_1$ defined as in Ref.~\cite{kravtsov2017non} satisfies $D_1=1-\kappa W^4<1$, as was also obtained numerically in that paper.

 Finally, we have revisited the derivation of the critical disorder in the limit of large disorder (equivalently, of large connectivity of the lattice), showing explicitly that the forward approximation fails to correctly capture the spectrum of the integral operator because it neglects relevant correlations between consecutive propagators along a path. We notice that for single particle problems, the asymptotic formula for $W_c$ has been shown to be in very good agreement with the numerical results, for dimensions $d\gtrsim 4$~\cite{tarquini2017critical}; therefore, it would be desirable to extend our analytic progress on the integral equation to the analytic treatment of MBL.

A natural continuation of this work would be the systematic derivation of a perturbative expansion for the eigenvalue $\lambda(s)$ of the integral kernel $K_{s}$, given by
\begin{equation}
    \lambda(s)= 1-s \, \theta_{\rm typ}+ O(s^2).
\end{equation}
This expansion is convergent because the spectrum of $K_{s=0}$ is gapped. In the case of a Cauchy distribution of disorder, this can be worked out using the explicit expressions for the eigenstates and for the spectrum of the operator $K_{s=0}$ reported in this work. In particular, in the large disorder limit it should be possible to identify which contributions in the expansion have to be retained  to recover the asymptotic correct behavior at $s=1/2$; this should give indications on how to properly account for the self-energy contributions in more conventional path expansions in the localized phase, in both tree-like and finite dimensional  lattices. We leave these calculations for future work  \cite{WorkInProgress}.

\section*{Acknowledgments}
The authors would like to thank V. Kravtsov, L.Ioffe, A.Mirlin and K.Tikhonov for interesting discussions.
FP thanks G. Lemari\'e for discussions and acknowledges the support of the project THERMOLOC ANR-16-CE30-0023-02 of the French National Research Agency (ANR). VR acknowledges the support of the Simons Foundation collaboration Cracking the Glass Problem (No. 454935 to G. Biroli).
SP is partially supported by Istituto Nazionale di Fisica Nucleare (INFN) through the project ``QUANTUM".
This project has received funding from the European Research Council  (ERC)  under  the  European  Union's  Horizon  2020  research  and  innovation programme (grant agreement 694925).\\

\appendix

\section{Eigenfunctions of the Cauchy Kernel at s=0 }\label{appendixA}

In this Appendix we derive the expressions for the right and left eigenfunctions of the Kernel Eq. (\ref{eq:Kcauchy}), given in Eqs.~\eref{eq:RightEigen}, \eref{eq:LeftEigen} in the main text. For the purpose of the derivation, it is convenient to consider the Kernel in Fourier space, which reads:
  \begin{equation}
  \tilde{K}(k',k)= \delta(k)- \sqrt{\frac{|k'|}{|k|}} J_1(2 \sqrt{|k| |k'|}) e^{- \overline{\alpha} |k|}\grafe{\theta(k) \theta(k')+ \theta(-k)\theta(-k')},
 \end{equation}
  where we assume $\theta(0)=0$, $\theta(x)\delta(x)=0$, and where $J_1 (\cdot)$ is a Bessel function of the first kind. We denote with $\tilde{\Psi}^{(n)}(k')$ and $\tilde{\Phi}^{(n)}(k)$ the left and right eigenvectors of the Kernel, respectively, with $n=0,1,...$ and with eigenvalues $\lambda_n=(-1)^n \alpha^{2n}$. We aim at showing that the left eigenvectors (up to normalization) are equal to:

  \begin{equation}
 \tilde{\Psi}^{(n)}(k') =
 \left\{ \begin{array}{@{\kern2.5pt}lL}
    \hfill \delta(k')  & for $n=0$.\\
    \hfill  - \frac{2 \delta(k') }{(1+ \alpha^2)^{n}}+  \frac{e^{-\frac{|k'|}{\alpha}}}{\alpha} S_n\quadre{\theta(k')+ \theta(-k')}\ & for $n \geq 1$.
    \end{array}\right.
 \end{equation}
where
\begin{equation}
S_n=\sum_{m=0}^{n-1} c_m^{(n)} L_m^{(1)}\tonde{\frac{|k'|}{\alpha}},
\end{equation}

$L_m^{(1)}(\cdot)$ are generalized Laguerre polynomial, and
\begin{equation}
c_m^{(n)}=  \tonde{\frac{-\alpha^2}{1+ \alpha^2}}^{n-1-m}{n \choose m+1}.
\end{equation}

For $n=0$, the eigenvalue equation is trivially satisfied by $\delta(k')$. For $n =1,2,\cdots$, the eigenvalue equation reads:
 \begin{equation}
  I_1(k)+ I_2(k)= \lambda_n \tilde{\Psi}^{(n)}(k),
 \end{equation}
where

  \begin{eqnarray}
&&I_1(k)=\delta(k) \int_{-\infty}^\infty dk' \tilde{\Psi}^{(n)}(k')\\
&=&\delta(k) \quadre{ - \frac{2}{(1+ \alpha^2)^{n}}+\frac{2}{\alpha}  \sum_{m=0}^{n-1} \quadre{\frac{-\alpha^2}{1+ \alpha^2}}^{n-1-m}{n \choose m+1} \int_0^\infty dk' \, e^{-\frac{|k'|}{\alpha}} L_m^{(1)}\tonde{\frac{|k'|}{\alpha}}}\nonumber\\
&=&\delta(k) \quadre{ - \frac{2}{(1+ \alpha^2)^{n}}+2  \frac{1-(-1)^{n} \alpha^{2n}}{(1+\alpha^2)^{n}}}= \delta(k) \quadre{- 2 \frac{(-1)^{n} \alpha^{2n}}{(1+\alpha^2)^{n}}},
\end{eqnarray}

while

\begin{equation}
 I_2(k)= -\grafe{\theta(k)+ \theta(-k)}\frac{e^{- \overline{\alpha} |k|}}{\alpha \sqrt{|k|}}  \sum_{m=0}^{n-1} \quadre{\frac{-\alpha^2}{1+ \alpha^2}}^{n-1-m}{n \choose m+1}  \mathcal{I}_m (|k|),
\end{equation}
and
\begin{equation}
 \mathcal{I}_m(|k|)=\int_{0}^{\infty} dk' \, e^{-\frac{k'}{\alpha}} L_m^{(1)}\tonde{ \frac{k'}{\alpha}}\sqrt{k'} J_1(2 \sqrt{|k| k'}).
\end{equation}
To compute this integral, we exploit the following identity, holding for any real $t$:
\begin{equation}
 J_\alpha(x)= \tonde{\frac{x}{2}}^\alpha \frac{e^{-t}}{\Gamma(\alpha+1)} \sum_{k=0}^\infty \frac{L_k^{(\alpha)}\tonde{\frac{x^2}{4 t}}}{{k+\alpha \choose k}} \frac{t^k}{k!}.
\end{equation}
Choosing $t=\alpha |k|$, we get

\begin{eqnarray}
 \mathcal{I}_m(|k|)&= \sqrt{|k|}\sum_{j=0}^\infty  e^{-\alpha |k|} \alpha^ j|k|^j \frac{1}{j! {j+1 \choose j}} \int_0^\infty dk' \, e^{-\frac{|k'|}{\alpha}} \,  L_j^{(1)}\tonde{\frac{k'}{\alpha}} L_m^{(1)}\tonde{\frac{k'}{\alpha}}\,k' \nonumber\\
 &=\sqrt{|k|}\sum_{j=0}^\infty  e^{-\alpha |k|} |k|^j \alpha^{j+2}\frac{1}{(j+1)!} \int_0^\infty du\, e^{-u} \,  L_j^{(1)}(u)  L_m^{(1)}(u)\, u \nonumber.
 \end{eqnarray}

Using the orthogonality condition for the generalized Laguerre polynomials:
\begin{equation}
 \int_0^\infty \, e^{-u} \,  L_j^{(1)}(u)  L_m^{(1)}(u)\, u=(m+1)\delta_{m,j},
\end{equation}
we have

\begin{eqnarray}
 \mathcal{I}_m(|k|)= \sqrt{|k|} e^{-\alpha |k|} |k|^m \alpha^{m+2} \frac{1}{m!}.
 \end{eqnarray}

 Now,
 \begin{equation}
  x^m= \sum_{l=0}^m \frac{(-1)^l m!}{(m-l)!}\frac{\Gamma(m+\beta+1)}{\Gamma(l+\beta+1)} L^{(\beta)}_l(x)
 \end{equation}
and thus

\begin{eqnarray}
 \mathcal{I}_m(|k|)= \sqrt{|k|} e^{-\alpha |k|}\alpha^{2m+2} \frac{1}{m!} \sum_{l=0}^m \frac{(-1)^l m!}{(m-l)!}\frac{(m+1)!}{(l+1)!} L^{(1)}_l\tonde{\frac{|k|}{\alpha}} .
 \end{eqnarray}

 This implies that
\begin{equation}
 I_2(k)= -\grafe{\theta(k)+ \theta(-k)}\frac{e^{- \tonde{\overline{\alpha}+\alpha} |k|}}{\alpha} \sum_{l=0}^{n-1} \frac{(-1)^l n!}{(l+1)!} S_l\, L^{(1)}_l\tonde{\frac{|k|}{\alpha}},
\end{equation}
 where
 \begin{eqnarray}
  S_l&= \sum_{m=l}^{n-1} \tonde{\frac{-\alpha^2}{1+ \alpha^2}}^{n-1-m}  \frac{\alpha^{2m+2}}{(n-1-m)!} \frac{1}{(m-l)!}\nonumber \\
  &= \tonde{\frac{\alpha^2}{1+\alpha^2}}^{n-l-1}\frac{1}{ (n-1-l)!} \alpha^{2n},
 \end{eqnarray}
so that
\begin{equation}
 I_2(k)=(-1)^{n} \alpha^{2n}\grafe{\theta(k)+ \theta(-k)}\frac{e^{- \tonde{\overline{\alpha}+\alpha} |k|}}{\alpha} \sum_{m=0}^{n-1} \frac{(-\alpha^2)^{n-1-m} }{(1+ \alpha^2)^{n-1-m}}{n \choose m+1}\, L^{(1)}_m \tonde{\frac{|k|}{\alpha}},
\end{equation}
 meaning that the eigenvalue equation is satisfied, given that $\alpha + \overline{\alpha}=\alpha^{-1}$.\\
 Following the same procedure, one can show that the right eigenvectors read (up to normalization):
  \begin{equation}
 \tilde{\Phi}^{(n)}(k') =
 \left\{ \begin{array}{@{\kern2.5pt}lL}
    \hfill e^{-\alpha |k|}  & for $n=0$.\\
    \hfill  e^{-\alpha |k|} \sum_{m=0}^{n-1} c_m^{(n)} \, |k| \, L_m^{(1)}\tonde{\frac{|k|}{\alpha}} \ & for $n \geq 1$.
    \end{array}\right.
\end{equation}
with the same coefficients $c_m^{(n)}$ defined above. \\
The expressions in Eqs.\ \eref{eq:RightEigen}, \eref{eq:LeftEigen} are obtained by performing the inverse Fourier transform. For the left eigenvectors we get:

\begin{eqnarray}
 \Psi^{(n)}(y)&= \int_{-\infty}^\infty dk'\, e^{i k' y} \tilde{\Psi}^{(n)}(k')\\
 &=- \frac{2}{(1+ \alpha^2)^{n}}+ \frac{2}{\alpha} \sum_{m=0}^{n-1} c_m^{(n)} \int_0^\infty dk' \, \cos( k' y) e^{-\frac{k'}{\alpha}} L_m^{(1)}\tonde{\frac{k'}{\alpha}}.
 \end{eqnarray}

We now exploit the generating function of the generalized Laguerre polynomials:
\begin{equation}
 \sum_{n=0}^\infty t^n \, L_n^{(1)}(x)= \frac{1}{(1-t)^2} e^{-\frac{t }{1-t} x},
\label{eq:LaguerreGF}
\end{equation}
so that
\begin{equation}
 L_n^{(1)}(x)=\frac{1}{n!} \lim_{t \to 0} \frac{d^n}{dt^n} \quadre{\frac{1}{(1-t)^2} e^{-\frac{t }{1-t} x}},
\end{equation}
to write

\begin{eqnarray}
&\int_0^\infty dk' \, \cos \tonde{ k' y} e^{-\frac{k'}{\alpha}} L_n^{(1)}\tonde{\frac{k'}{\alpha}} \nonumber\\
&=\frac{1}{n!}\lim_{t \to 0} \frac{d^n}{dt^n} \quadre{ \frac{1}{(1-t)^2} \int_0^\infty dk' \, \cos \tonde{ k' y}  e^{-\frac{1}{1-t} \frac{k'}{\alpha}}}\nonumber\\
&=\frac{\alpha}{n!}\lim_{t \to 0} \frac{d^n}{dt^n} \quadre{ \frac{1}{(1-t)}\frac{1}{\alpha^2(1-t)^2 y^2+ 1}},
\end{eqnarray}

and thus
\begin{eqnarray}\label{eq:LeftESpace}
\Psi^{(n)}(y)=&- \frac{2 }{(1+ \alpha^2)^{n}}+ 2 \sum_{m=0}^{n-1} \quadre{\frac{-\alpha^2}{1+ \alpha^2}}^{n-1-m}{n \choose m+1} \times \nonumber\\
&\frac{1}{m!}\lim_{t \to 0} \frac{d^m}{dt^m} \quadre{ \frac{1}{(1-t)}\frac{1}{\alpha^2(1-t)^2 y^2+ 1}},
\end{eqnarray}
which is Eq. \eref{eq:LeftEigen} in the main text, up to a normalization factor.

For the right eigenvector, we need to evaluate the following integral:
\begin{eqnarray}
 \Phi^{(n)}(x)= \int_{-\infty}^\infty dk\, e^{-i k x} \tilde{\Phi}^{(n)}(k)=2 \sum_{m=0}^{n-1} c_m^{(n)} \int_0^\infty dk \, \cos( k x)\, k\, e^{-\alpha k} L_m^{(1)}\tonde{\frac{k}{\alpha}}\nonumber.
\end{eqnarray}
Exploiting once more the generating function \eref{eq:LaguerreGF}, we get
\begin{eqnarray}
&\int_0^\infty dk \, \cos( k y)\, k\, e^{-\alpha k} L_m^{(1)}\tonde{\frac{k}{\alpha}} \nonumber\\
&=\frac{1}{m!}\lim_{t \to 0} \frac{d^m}{dt^m} \quadre{ \frac{1}{(1-t)^2} \int_0^\infty dk \, \cos \tonde{ k x} \, k\, e^{-\frac{\alpha^2+(1-\alpha^2)t}{\alpha(1-t)}  k}} \nonumber\\
&=\frac{\alpha^2}{m!}\lim_{t \to 0} \frac{d^m}{dt^m} \quadre{\frac{-\alpha^2(1-t)^2 x^2+ (\alpha^2+t(1-\alpha^2))^2}{\quadre{\alpha^2(1-t)^2 x^2+ (\alpha^2+t(1-\alpha^2))^2}^2}}
\end{eqnarray}
and so
\begin{eqnarray}
\Phi^{(n)}(x)&=& 2\alpha^2 \sum_{m=0}^{n-1} \quadre{\frac{-\alpha^2}{1+ \alpha^2}}^{n-1-m}{n \choose m+1} \times \nonumber \\
&&\frac{1}{m!}\lim_{t \to 0} \frac{d^m}{dt^m} \quadre{\frac{-\alpha^2(1-t)^2 x^2+ (\alpha^2+t(1-\alpha^2))^2}{\quadre{\alpha^2(1-t)^2 x^2+ (\alpha^2+t(1-\alpha^2))^2}^2}},
\end{eqnarray}
which is Eq. \eref{eq:RightEigen} in the main text, up to a normalization factor.\\

\newpage

\bibliographystyle{unsrt.bst}

\bibliography{MBLbib.bib}

\end{document}